\documentclass[12pt]{article}%
\usepackage{tabularx}%
\usepackage{caption}%
\usepackage{subcaption}%

\usepackage[utf8]{inputenc}%
\usepackage{placeins}%

\usepackage{lineno}%
\usepackage{multirow}%
\usepackage{makecell}%
\usepackage{siunitx}%
\usepackage{adjustbox}%
\usepackage{pifont}%
\usepackage{color}%
\usepackage{nicematrix}%
\usepackage{comment}%
\usepackage[hidelinks,hyperfootnotes=false]{hyperref}%
\usepackage{booktabs}%
\usepackage{threeparttable}%
\usepackage{apacite}%
\usepackage{parskip}%
\usepackage{array}%
\newcolumntype{R}{>{\raggedleft\arraybackslash}X}%
\newcolumntype{L}{>{\raggedright\arraybackslash}X}%
\newcolumntype{Y}{>{\centering\arraybackslash}X}%
\newcommand{\ok}[1]{{\leavevmode\color{black}#1}}%
\newcommand{\notouch}[1]{{\leavevmode\color{black}#1}}%
\usepackage{tikz}%
\usetikzlibrary{backgrounds,positioning,shapes.geometric, arrows}%

\usepackage{setspace}%
\usepackage[letterpaper,top=1in,bottom=1in,right=0.8in,left=.9in]{geometry}
\usepackage{newtxtext,newtxmath} 
\usepackage[T1]{fontenc}
\usepackage{microtype}

\makeatletter
\renewcommand\maketitle{%
  \vspace*{-0.5\baselineskip}
  {\Large\bfseries\@title\par}
  \vspace{1.0\baselineskip}
  {\normalsize\@author\par}
  \vspace{1.2\baselineskip}
}
\makeatother

\title{{\noindent}%
How Ten Publishers Retract Research
}%

\author{%
	\noindent
	Jonas Oppenlaender 
	\\
University of Oulu
	\\
Oulu, Finland
	\\
jonas.oppenlaender@oulu.fi
}%

\begin{document}%
\maketitle%





\vspace{-1\baselineskip}

\subsection*{Abstract}
Retractions are the primary mechanism for correcting the scholarly record, yet publishers differ markedly in how they use them.
We present a bibliometric analysis of 46,087 retractions across 10 major publishers using data from the Retraction Watch database (1997--2026), examining retraction rates, reasons, temporal trends, and geographic distributions, among other dimensions.
Normalized retraction rates vary by two orders of magnitude, from Elsevier's 3.97 per 10,000 publications to Hindawi's 320.02.
China-affiliated authors account for the largest share of retractions at every publisher.
%
Retraction lags and reason profiles also vary widely across publishers.
Among the ten publishers, ACM is an outlier in its retraction profile.
ACM's normalized rate is mid-range (5.65), yet 98.3\% of its 354 retractions 
are related to one incident.
Seven of the ten most common global retraction reasons (including misconduct, plagiarism, and data concerns) are entirely absent from ACM's
record.
ACM's first retraction dates to 2020, despite a catalog dating to 1997.
ACM self-describes its retraction threshold as ``extremely high.''
%
We discuss this threshold in relation to the COPE retraction guidelines and the implications of ACM's non-public dark archive of removed works.%


\vspace{1\baselineskip}

\noindent
\textbf{Keywords:}
  retractions,
  research integrity,
  bibliometrics,
  scholarly publishing,
  digital libraries,
  publication ethics
%
%
%

\clearpage

\section{Introduction}%
\label{sec:introduction}%
\notouch{%
Retractions are the primary mechanism by which the scholarly record is formally corrected.
When a published article is found to contain serious errors, fabricated data, or evidence of misconduct, a retraction notice signals to the research community that the work should no longer be relied upon.
In 2023, over 10,000 papers were retracted in scientific publishing~\cite{d41586-023-03974-8.pdf}, although this figure is 
regarded as only the tip of the iceberg~\cite{Oransky}.
The scale of the problem is not merely an accounting concern.
The consequences extend beyond the retracted articles themselves: retracted findings continue to circulate in citations, policy documents, and downstream research~\cite{qss_a_00303.pdf, qss_a_00243.pdf, e031909.full.pdf}.
}%

\notouch{%
But retractions do not happen uniformly. The publishers who steward the scholarly record differ in how they detect, adjudicate, and formalize corrections.
Publishers differ in editorial structures, disciplinary cultures, and institutional norms.
\citeA{cetina2007.pdf} characterizes these differences as distinct epistemic cultures governing how knowledge is produced, validated, and corrected.
These differences shape how misconduct is detected, adjudicated, and ultimately acted upon.
At the same time, the systemic pressures that give rise to misconduct -- the publish-or-perish culture~\cite{edwards-roy-2017-academic-research-in-the-21st-century-maintaining-scientific-integrity-in-a-climate-of-perverse.pdf, tijdink2014.pdf}, the operations of paper mills~\cite{else2021.pdf, zhang-wang-2024-research-misconduct-in-china-towards-an-institutional-analysis.pdf,paper-mill-challenges,Scancare087581}, and institutional reward structures~\cite{Elite} -- are not confined to specific publishers or disciplines.
A publisher's retraction record therefore reflects not only the prevalence of misconduct in its corpus but also the publisher's willingness and capacity to act on it.
Cross-publisher comparison can reveal differences in policy and practice, not merely differences in misconduct rates.
}%

\notouch{%
In this study, we analyze 46,087 retractions across ten publishers: Hindawi, IEEE, Springer Nature, Elsevier, Wiley, Taylor~\&~Francis, SAGE, IOS Press, PLoS, and ACM.
Data were collected from the Retraction Watch database~\cite{retractionwatch} 
 and Dimensions \cite{Hook2018Dimensions}.
Our analyses encompass retraction rates, reasons, temporal trends, geographic distributions, and other dimensions detailed in Section~\ref{sec:analyses}.
}%

\notouch{%
The cross-publisher comparison reveals substantial variation, with no single retraction profile constituting a ``normal'' baseline.
Publishers differ in volume (from 354 to over 11,000 retractions), timing (median lags ranging from weeks to years), reason composition (near-monolithic to broadly distributed), and geographic distribution. One geographic pattern is consistent across all ten publishers: China-affiliated authors account for the largest share of retractions at every publisher examined, reflecting systemic pressures~\cite{zhang-wang-2024-research-misconduct-in-china-towards-an-institutional-analysis.pdf, Elite} not specific to any one venue.
}%

\notouch{%
Within this variation, ACM is a compositional outlier. Its normalized retraction rate is mid-range, comparable to publishers of similar size. The rate is not unusual. The \emph{composition} is: ACM's record is dominated by a single retraction reason and a single event~\cite{more-than-300-at-once}. ACM also maintains a non-public ``dark archive'' of removed works~\cite{ACM-POLICY}, the contents of which the scholarly community cannot independently verify.
Because this compositional profile and the policy choices behind it raise questions about threshold-setting, detection, and transparency, we discuss ACM's record in detail alongside the broader cross-publisher patterns.
}%

\notouch{%
This study makes three contributions.
First, it provides a cross-publisher retraction comparison across ten publishers and nine dimensions, documenting variation in volume, timing, 
reason diversity, geographic distribution, and repeat-author rates, among others. No single publisher profile constitutes a baseline; each reflects distinct editorial cultures and policy choices.
Second, it identifies ACM as a compositional outlier whose retraction record is dominated by a single reason category from a single event, a pattern that persists after controlling for publication volume and that cannot be attributed to sample size (permutation test, $p < 0.001$).
Third, it examines the gap between ACM's self-described ``extremely high'' retraction threshold and the COPE guidelines ACM has adopted for a subset of its journals, and discusses the transparency implications of a non-public dark archive whose contents the scholarly community cannot independently verify.
}%

\notouch{%
The sections that follow trace these patterns from context through data to consequence. We review the retraction landscape and the policies that govern it (Section~\ref{sec:relatedwork}), describe our dataset and methods (Section~\ref{sec:method}), present cross-publisher findings (Section~\ref{sec:results}), and discuss implications for retraction practice and research integrity (Sections~\ref{sec:discussion} and~\ref{sec:conclusion}).
}%

\section{Background and Related Work}%
\label{sec:relatedwork}%

\notouch{%
Situating the cross-publisher comparison requires context: how retractions function, what drives them, and how publisher policies shape the resulting record.
We first define the corrective actions available to publishers (Section~\ref{sec:rel:retractions}), then review the scale and growth of retractions (Section~\ref{sec:rel:scale}), the drivers behind them (Section~\ref{sec:rel:drivers}), and their consequences (Section~\ref{sec:rel:consequences}).%
}%

\subsection{Retractions and the Scholarly Record}%
\label{sec:rel:retractions}%
\notouch{%
Four corrective actions are available to authors and publishers for maintaining the integrity of the published record: withdrawal, correction, retraction, and removal.
They differ in severity, timing, and who initiates them.
}%

\notouch{%
\textit{Withdrawal} is an author's request to remove work before publication.
The publication process is halted and the withdrawn work does not enter the academic record.
\textit{Corrections} address minor flaws or errors in published works, typically implemented as an erratum or corrigendum appended to the original.
\textit{Retraction} notifies readers that a publication should not be relied upon.
A retraction is warranted when significant flaws invalidate an article's findings, or when there is evidence of misconduct such as data fabrication or plagiarism.
}%
\notouch{%
Retracted articles typically remain on the publisher's website; instead of deletion, a retraction notice explains the reason and maintains transparency.
The notice usually includes a title change, an explanation, and additional markings such as watermarks in the PDF.
\textit{Removal} is the most severe response.
While a retracted work remains publicly accessible in a marked form, a removed work is eliminated from the publisher's online repository entirely.
The publisher may keep a private copy for reference, but the public cannot access it.
}%

\notouch{%
COPE has published guidelines that serve as the de facto standard for retraction practices~\cite{retraction-guidelines-cope.pdf}.
COPE is a membership organization: member publishers commit to following its guidelines on publication ethics, including retraction.
	Of the ten publishers examined in this study, nine are confirmed COPE members.\footnote{See \url{https://members.publicationethics.org/members}.}
	Hindawi is not listed in the COPE directory, but was acquired by Wiley during the study period.
Publishers vary in how they apply COPE guidelines, and membership does not mean that all of a publisher's publications are covered.
	ACM, for instance, is a COPE member for 48 journals, not its full publication portfolio.
}%

\notouch{%
Retractions and removals carry financial costs for the organizations that funded the original work~\cite{elife-02956-v1.pdf}.
Prior investments in data collection, personnel, and related activities become sunk costs, and funding bodies may face additional commitments to correct the record or support follow-up studies.
Retractions can also be a harrowing experience for those who have been plagiarized.
Cases have been reported in which researchers were unsuccessful in getting their plagiarized works retracted from predatory publishers~\cite{disturbing-experience}.
This highlights the importance of clear and enforceable retraction policies.
}%

\subsection{The Scale and Growth of Retractions}%
\label{sec:rel:scale}%

\notouch{%
The number of retracted publications has grown over the past two decades.
In 2023, more than 10{,}000 research papers were retracted which is a new annual record~\cite{d41586-023-03974-8.pdf}.
However, this figure is regarded as the tip of the iceberg~\cite{Oransky}.
For instance, \citeA{Scancare087581} found a large increase of potentially fraudulent papers in cancer research, stemming, to a large degree, from Chinese authors.
}%

\notouch{%
The Retraction Watch database~\cite{retractionwatch} is the most comprehensive public catalog of retraction notices. The database was acquired by Crossref in 2023. With this acquisition, retraction metadata was made openly available to the scientific community~\cite{Crossref}.
This growing data availability has enabled large-scale bibliometric research on retractions, making it an active focus area of meta-research~\cite{MetaResearch} and the ``science of science''~\cite{science.aao0185.pdf}.
Retraction studies span multiple disciplines, including
information science~\cite{3576840.3578281},
computer science~\cite{pone.0285383.pdf},
psychology~\cite{yang2020.pdf},
and the humanities~\cite{qss_a_00222.pdf}.
The present study 
compares retraction patterns across ten major academic publishers.
}%

\subsection{Drivers of Retractions}%
\label{sec:rel:drivers}%

\notouch{%
Retractions arise from a mix of systemic pressures, individual misconduct, and technological change.
A primary systemic factor is the publish-or-perish culture.
``Perverse'' incentive structures that reward quantity over quality have been shown to erode scientific integrity~\cite{edwards-roy-2017-academic-research-in-the-21st-century-maintaining-scientific-integrity-in-a-climate-of-perverse.pdf}, and publication pressure has been linked to a higher willingness to engage in questionable research practices~\cite{tijdink2014.pdf}.
}%

\notouch{%
A growing concern is the rise of paper mills, organizations that produce and sell fabricated manuscripts to researchers seeking to inflate their publication records~\cite{else2021.pdf}.
Paper mills have been linked to large volumes of retractions, particularly in certain national contexts.
Research misconduct in China has received substantial scholarly attention~\cite{zhang-wang-2024-research-misconduct-in-china-towards-an-institutional-analysis.pdf,science.342.6162.1035}, with reports indicating that even elite researchers have felt compelled to engage in misconduct~\cite{Elite}.
Misconduct has also been documented in hospital-affiliated research settings~\cite{Research-misconduct-in-hospitals-is-spreading-A-bibliometric-analysis-of-retracted-papers-from-Chinese-universityaffiliated-hospitals.pdf}.
These patterns reflect broader institutional and cultural factors that vary across epistemic cultures~\cite{cetina2007.pdf}.
}%

\notouch{%
Technology has introduced new vectors.
Large language models have raised concerns about AI-generated scientific content~\cite{2406.07016.pdf}, including GPT-fabricated papers appearing on Google Scholar~\cite{haider_gpt_fabricated_scientific_papers_20240903.pdf} and ``tortured phrases,'' machine-generated paraphrases designed to evade plagiarism detection~\cite{2107.06751.pdf}.
Image manipulation in scientific figures is another documented concern~\cite{zlabinger2017.pdf, FEBS}.
High-profile cases of individual misconduct, such as that of former Harvard Business School Professor Francesca Gino~\cite{gino}, have further increased public awareness of research integrity issues.
}%

\subsection{Consequences of Retracted Research}%
\label{sec:rel:consequences}%

\notouch{%
Retracted research continues to circulate in the scientific literature long after retraction~\cite{qss_a_00303.pdf, s11192-022-04321-w.pdf, qss_a_00222.pdf, pone.0285383.pdf}.
This persistence threatens the integrity of downstream research, as subsequent studies may build upon flawed findings without awareness of the retraction~\cite{e031909.full.pdf}.
The problem extends beyond academia: retracted research has been found in public policy documents~\cite{qss_a_00243.pdf} and on social media platforms, where it may cater to prior beliefs and fuel misinformation~\cite{3637335.pdf}.
}%

\notouch{%
The COVID-19 pandemic illustrated the urgency of these concerns.
The rapid pace of publication was accompanied by a notable increase in retracted pharmacotherapy-related articles~\cite{el-menyar-et-al-2021-publications-and-retracted-articles-of-covid-19-pharmacotherapy-related-research-a-systematic.pdf}. This pattern illustrates the risks of a publishing ecosystem in which speed can take precedence over rigor.
The continued circulation of retracted research also poses challenges for information retrieval systems.
\citeA{3576840.3578281} examined how retracted articles surface in search results and highlighted the need for digital libraries to more prominently flag retracted content.
}%

\notouch{%
How publishers detect, disclose, and label retracted work determines whether downstream researchers, policymakers, and retrieval systems can respond to flawed findings.
Yet few studies have compared how individual publishers handle retractions.
}%

\section{Method}%
\label{sec:method}%

\notouch{%
Our aim is to compare publishers' retraction cultures and norms as reflected in their retraction records. To do so, we assembled a cross-publisher dataset from the Retraction Watch database (Section~\ref{sec:data-collection}), consolidated its \ok{111} reason codes into broader categories suitable for comparison (Section~\ref{sec:reason-consolidation}), normalized retraction counts by publication volume using external data sources (Section~\ref{sec:normalization}), and applied a series of descriptive and inferential analyses across nine dimensions (Section~\ref{sec:analyses}). Section~\ref{sec:limitations} discusses the limitations of this approach.
}%

\subsection{Data Collection and Pre-Processing}%
\label{sec:data-collection}%

\notouch{%
We obtained retraction data from the Retraction Watch database~\cite{retractionwatch}, a comprehensive, freely accessible database of retracted scholarly works maintained by Crossref~\cite{Crossref}.
We used Retraction Watch rather than querying publisher repositories directly for three reasons.
First, Retraction Watch provides a unified interface across multiple publishers, minimizing inconsistencies from manual queries against individual digital libraries.
Second, because ACM maintains a ``dark archive'' of removed articles~\cite{ACM-POLICY}, Retraction Watch may capture records that are no longer publicly accessible in the ACM-DL\@.
Third, Retraction Watch pre-categorizes entries with structured reason codes, providing a ready-made framework for comparative analysis.
}%

\begin{table*}[htbp]
	\centering
	\small
	\caption{Retraction nature by publisher. Each entry in the Retraction Watch database is classified as a retraction, expression of concern, correction, or reinstatement.}
	\label{tab:retraction-nature}
	\begin{tabular}{l r r r r r r r r}
		\toprule
		& \multicolumn{2}{c}{Retraction} & \multicolumn{2}{c}{Expression of} & \multicolumn{2}{c}{Correction} & \multicolumn{2}{c}{Reinstatement} \\
		& \multicolumn{2}{c}{} & \multicolumn{2}{c}{Concern} & \multicolumn{2}{c}{} & \multicolumn{2}{c}{} \\
		\cmidrule(lr){2-3} \cmidrule(lr){4-5} \cmidrule(lr){6-7} \cmidrule(lr){8-9}
		Publisher & N & \% & N & \% & N & \% & N & \% \\
		\midrule
		ACM & 354 & 99.7 & 1 & 0.3 & 0 & 0.0 & 0 & 0.0 \\
		IEEE & 10,094 & 100.0 & 1 & 0.0 & 1 & 0.0 & 0 & 0.0 \\
		Elsevier & 6,520 & 87.7 & 575 & 7.7 & 257 & 3.5 & 86 & 1.2 \\
		Springer Nature & 7,534 & 88.6 & 726 & 8.5 & 238 & 2.8 & 2 & 0.0 \\
		Wiley & 3,817 & 95.0 & 87 & 2.2 & 114 & 2.8 & 0 & 0.0 \\
		Taylor \& Francis & 2,314 & 89.8 & 201 & 7.8 & 52 & 2.0 & 10 & 0.4 \\
		SAGE & 1,217 & 71.2 & 467 & 27.3 & 25 & 1.5 & 0 & 0.0 \\
		Hindawi & 11,486 & 99.7 & 22 & 0.2 & 16 & 0.1 & 0 & 0.0 \\
		IOS Press & 1,671 & 99.9 & 1 & 0.1 & 1 & 0.1 & 0 & 0.0 \\
		PLoS & 1,080 & 78.4 & 212 & 15.4 & 86 & 6.2 & 0 & 0.0 \\
		\midrule
		\textbf{Total} & 46,087 & 93.5 & 2,293 & 4.7 & 790 & 1.6 & 98 & 0.2 \\
		\bottomrule
	\end{tabular}
\end{table*}

\notouch{%
We downloaded the full Retraction Watch database on February~14, 2026.
From this dataset, we selected the nine publishers with the highest number of retraction entries, after consolidating sub-brands under their parent companies (e.g., Springer, Springer -- Nature Publishing Group, and Springer -- Biomed Central under Springer Nature; Cell Press under Elsevier; Dove Press under Taylor~\&~Francis).
We additionally included ACM, for a total of ten publishers. ACM was included because it is a major professional-society publisher that has not previously been examined in cross-publisher retraction studies.
Two publishers in this set changed ownership during the study period: Hindawi was acquired by Wiley in January 2023, and IOS Press was acquired by SAGE Publications in November 2023. We retained both as separate entities because they operated independently for the majority of the period covered by the retraction records.
The resulting dataset comprises 49,268 entries spanning 1997 to early 2026.
The start year coincides with the launch of ACM's Digital Library, ensuring comparable temporal coverage across all ten publishers.
Table~\ref{tab:retraction-nature} shows the breakdown by retraction nature: 46,087 (93.5\%) are retractions, 2,293 (4.7\%) expressions of concern, 790 (1.6\%) corrections, and 98 (0.2\%) reinstatements.
Because expressions of concern, corrections, and reinstatements are operationally distinct from retractions, all subsequent analyses are restricted to the 46,087 retractions.
}%

%

%


\subsection{Reason Consolidation}%
\label{sec:reason-consolidation}%

\notouch{%
The Retraction Watch database assigns one or more reason codes to each entry using a controlled vocabulary of \ok{111} distinct codes.
Multiple reasons may be associated with a single entry (semicolon-delimited).
To facilitate comparative analysis, we consolidated semantically related reason codes into broader categories using substring-based substitution rules.
The mapping contains 51 rules that rewrite fine-grained reason codes into 17 target categories.
For example, ``Plagiarism of Text,'' ``Plagiarism of Article,'' ``Plagiarism of Image,'' ``Euphemisms for Plagiarism,'' ``Copyright Claims,'' and ``Taken from Dissertation/Thesis'' all map to \emph{Plagiarism}.
Of the 111 original reason codes, 77 are not covered by any substitution rule and pass through unchanged (e.g., ``Compromised Peer Review,'' which is already sufficiently specific).
After consolidation, 95 unique reason codes remain.
The full mapping is provided in Table~\ref{tab:reason-consolidation}.
}%

\notouch{%
To assess robustness of the consolidation, we tested three alternative mappings that reassign contested codes (e.g., Conflict of Interest from Authorship to Misconduct, Copyright Claims from Plagiarism to Other).
Under these alternatives, per-publisher reason distributions remain stable: top-3 rankings are unchanged for all publishers and the maximum percentage-point shift stays below 2.6~pp for eight of ten publishers (Table~\ref{tab:sensitivity-reasons}).
}%

\notouch{%
For comparative analyses of retraction reasons across publishers, we excluded six meta-reasons that describe procedural or administrative status rather than substantive retraction causes: ``Investigation by Journal/Publisher,'' ``Information,'' ``Date of Article and/or Notice Unknown,'' ``Notice,'' ``Upgrade/Update of Prior Notice(s),'' and ``Notice -- Unable to Access via current resources.''
}%

\notouch{%
For visualization, figures display only the ten most frequent consolidated reason codes across all publishers.
These reasons were selected by counting each code's total
mentions across all 46,087 retractions (after meta-reason exclusion), ranking by frequency, and retaining the ten highest-ranked codes. The
same ten codes are used in Figures~\ref{fig:reasons-by-publisher} and~\ref{fig:heatmap-panel}. The full set of 87 substantive codes
underlies all statistical analyses.
}%

\subsection{Normalization and External Data Sources}%
\label{sec:normalization}%

\notouch{%
To compare retraction rates across publishers of varying size, we normalized retraction counts by total publication volume.
We queried the Dimensions Analytics API~\cite{Hook2018Dimensions} for total publication counts per publisher over the period 1997--2026, filtering to research articles (articles and conference proceedings).
The retraction data were filtered analogously to include only research-type entries (i.e., excluding retraction notices, commentaries, editorials, and letters).
Normalized retraction rates are expressed as retractions per 10,000 publications.
For this analysis, sub-brands were merged into their parent companies, consistent with Section~\ref{sec:data-collection}.
To validate these counts, we cross-checked the Dimensions totals against the Crossref REST API, querying each publisher's member record for journal articles and proceedings articles over the same period.
The two sources agree within 9\% for nine of ten publishers (Pearson~$r = 0.9985$). The exception is Hindawi, where post-acquisition DOI reassignment to Wiley's Crossref member record accounts for the discrepancy (see Appendix~\ref{appendix:crossref-validation}).
IOS Press's total publication count includes book series chapters (53{,}762) in addition to journal articles (4{,}032), because the majority of IOS Press content is published in book series that Dimensions classifies as chapters, while the Retraction Watch entries are labeled as research articles.
}%

\notouch{%
For author disambiguation in the repeat-author analysis, we queried the Dimensions API by DOI to retrieve \texttt{researcher\_id} fields, persistent identifiers assigned by Dimensions to disambiguate individual researchers.
For authors lacking a \texttt{researcher\_id}, we used author name combined with institutional affiliation as a composite key. Institution data came from two sources: per-author affiliations returned by the Dimensions API (available for 98.6\% of author records) and the entry-level Institution field in Retraction Watch (covering 99.99\% of entries).
Of the 148,508 unique author identifiers in our dataset, 111,805 (75.3\%) were disambiguated via a Dimensions \texttt{researcher\_id}, 36,661 (24.7\%) via name combined with institution, and 42 (0.03\%) via name only.
A sensitivity analysis comparing three disambiguation strategies (name-only, name+institution, and Dimensions-only) is provided in Appendix~\ref{appendix:sensitivity-authors}.
}%

\subsection{Analyses}%
\label{sec:analyses}%
\notouch{%
We conducted analyses across the following dimensions: descriptive statistics of retraction counts, temporal trends, and geographic distributions; normalized retraction rates; retraction reason distributions (using the consolidated categories from Section~\ref{sec:reason-consolidation}); retraction lag (time elapsed between original publication and retraction); country co-authorship patterns; repeat-author identification (using Dimensions' \texttt{researcher\_id} for disambiguation, with fallback to a composite ID);
paywall status; and article type distributions.
To quantify reason concentration, we computed the Herfindahl-Hirschman Index (HHI) and Shannon entropy on per-publisher reason-mention shares, tested each publisher's HHI against a null model of random sampling from the pooled distribution (permutation test, $B = 10{,}000$), and compared ACM's reason distribution to the remaining publishers using a chi-square goodness-of-fit test with Monte Carlo simulation ($B = 10{,}000$).
}%

\subsection{Limitations}%
\label{sec:limitations}%
\notouch{%
Several limitations should be considered.
First, Retraction Watch coverage may be incomplete; not all retractions issued by publishers are necessarily recorded in the database.
This limitation applies symmetrically: all ten publishers are measured through the same data source, and any under-counting would need to disproportionately affect one publisher to bias the comparison. Since Crossref acquired the Retraction Watch database in 2023~\cite{Crossref}, formally registered retraction DOIs are integrated into the dataset, reducing the likelihood of systematic omissions for publishers that register retractions with Crossref.
Second, ACM's ``dark archive''~\cite{ACM-POLICY} means that some retracted or removed content may not appear in any public database, including Retraction Watch, potentially leading to under-reporting of ACM retractions.
Of ACM's four corrective actions~\cite{ACM-POLICY}, withdrawal occurs before publication and corrections address minor errors. Neither falls within Retraction Watch's scope. The open question is whether ACM issues formal retractions or removals that are not registered with Crossref and therefore absent from the database.
Third, 0.03\% of authors in our dataset (42 of 148,508) were identified by name only, without institutional affiliation, which may conflate distinct individuals who share common names or fail to link name variants of the same individual. The sensitivity analysis in Appendix~\ref{appendix:sensitivity-authors} confirms that repeat-author counts are stable across disambiguation strategies.
Fourth, the consolidation of \ok{111} reason codes into broader categories involves subjective mapping decisions. A sensitivity analysis with three alternative groupings (see Table~\ref{tab:sensitivity-reasons}) shows that per-publisher distributions are largely stable, though individual codes such as Conflict of Interest (16.4\% of PLoS entries) can produce notable shifts when reassigned.
Fifth, the reason field in Retraction Watch varies in completeness across publishers. IEEE, for example, has 9,526 retractions tagged ``Notice -- Limited or No Information,'' the single largest reason category in the dataset. This means a substantial share of IEEE's retraction notices do not disclose the underlying cause. Cross-publisher comparisons of retraction reason distributions should be interpreted with this asymmetry in mind.
Sixth, this analysis relies entirely on publicly available data and published policies. ACM's retraction policy~\cite{ACM-POLICY} is the primary public document available for interpreting its retraction record. Internal deliberation processes, if any exist beyond what the policy describes, are not publicly documented.
Finally, this study is purely bibliometric. We do not infer intent, causality, or underlying misconduct beyond what is recorded in the Retraction Watch database. The absence of a retraction does not imply the absence of problems in the underlying work.
}%


\section{Results}%
\label{sec:results}%

\subsection{Dataset Overview}%
\label{sec:dataset-overview}%

\notouch{%
Table~\ref{tab:descriptive-raw-counts} summarizes the retraction counts per publisher.
Hindawi leads with 11,486 retractions (24.9\%), followed by IEEE (10,094; 21.9\%), Springer Nature (7,534; 16.3\%), and Elsevier (6,520; 14.1\%).
ACM contributes 354 retractions (0.8\%), the smallest count among the ten publishers.
}%

\begin{table}[!htbp]
\caption{Retraction counts and normalized retraction rates per publisher in the Retraction Watch database (retractions only; see Section~\ref{sec:data-collection}); total publication counts from the Dimensions database (1997--2026). Normalized rates are computed on research articles only (see Section~\ref{sec:normalization}). Sub-brands are merged into parent companies (see Section~\ref{sec:data-collection}).}
\label{tab:descriptive-raw-counts}
\centering
\small
\begin{tabular}{lrrrrr}
\toprule
Publisher & Retractions & \%  & Total Publications  & Rate per 10k & Paywalled (\%) \\
\midrule
Hindawi                       & 11,486  & 24.9\% & 358,879    & 320.02 & 0.0 \\
IEEE                          & 10,094  & 21.9\% & 5,703,250  & 17.70  & 5.8 \\
Springer Nature               &  7,534  & 16.3\% & 7,994,821  &  9.06  & 2.5 \\
Elsevier                      &  6,520  & 14.1\% & 15,686,751 &  3.97  & 1.2 \\
Wiley                         &  3,817  &  8.3\% & 6,071,179  &  6.21  & 4.5 \\
Taylor \& Francis             &  2,314  &  5.0\% & 3,534,113  &  6.50  & 1.8 \\
SAGE                          &  1,217  &  2.6\% & 2,179,130  &  5.46  & 3.2 \\
IOS Press                     &  1,671  &  3.6\% & 57,794     & 283.77 & 0.1 \\
PLoS                          &  1,080  &  2.3\% & 401,947    & 26.82  & 0.0 \\
ACM                           &    354  &  0.8\% & 626,110    &  5.65  & 0.0 \\
\midrule
\textbf{Total}                & \textbf{46,087}  &  & \textbf{42,613,974}  &  &  \\
\bottomrule
\end{tabular}
\end{table}

\notouch{%
To account for differences in publication volume, we normalized retraction counts by total publications per publisher using Dimensions API data for 1997--2026 
(see Table~\ref{tab:descriptive-raw-counts}). 
Hindawi and IOS Press have the highest normalized rates (320.02 and 283.77 per 10,000, respectively), reflecting mass retractions against relatively small publication bases.
Among the remaining publishers, PLoS (26.82) and IEEE (17.70) show the highest rates.
ACM's rate of 5.65 retractions per 10,000 publications places it eighth of ten, comparable to SAGE (5.46) and above only Elsevier (3.97).
Elsevier, despite the largest absolute publication volume (15.69 million), has the lowest retraction rate.
}%

\begin{figure*}[!htbp]
	\centering
	\includegraphics[width=\linewidth]{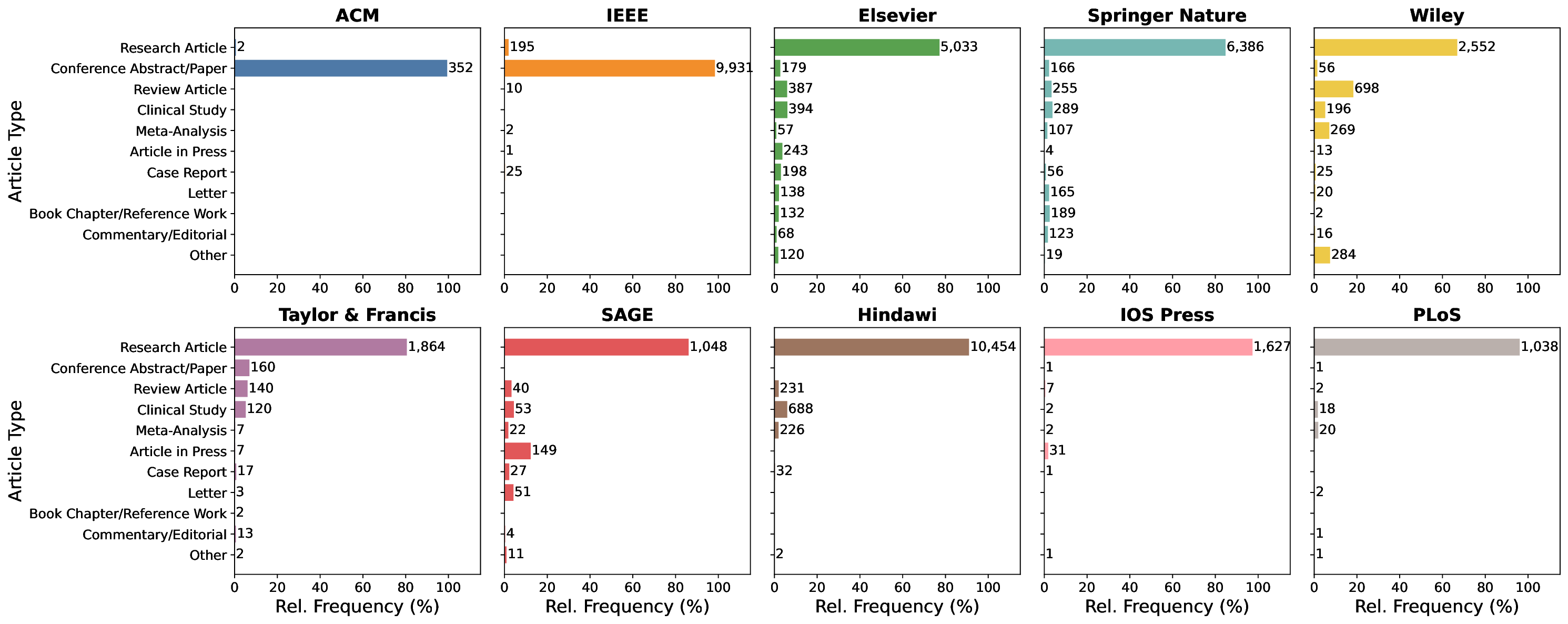}
	\caption{Distribution of article types per publisher. Bars show relative frequency (\%).
		Absolute counts are annotated at bar ends. Minor types are grouped as ``Other.''}
	\label{fig:article-types-by-publisher}
\end{figure*}

\notouch{%
Table~\ref{tab:descriptive-raw-counts} also reports the percentage of 
paywalled retractions.
Across all publishers, 2.4\% of retractions are paywalled.
IEEE has the highest paywall rate at 5.8\%, followed by Wiley (4.5\%) and SAGE (3.2\%).
Hindawi, PLoS, and ACM have 0\% paywall rates.
}%

\notouch{%
The dataset spans 23 article type labels assigned by Retraction Watch.
Across all publishers, Research Articles account for 65.5\% of retractions (30,199) and Conference Abstract/Paper for 23.5\% (10,846), followed by Clinical Studies (3.8\%), Review Articles (3.8\%), and Meta-Analyses (1.5\%) (see Figure~\ref{fig:article-types-by-publisher}).
A 
compositional divide separates ACM and IEEE from the rest:
ACM retractions are 99.4\% Conference Abstract/Paper (352 of 354), and IEEE retractions are 98.4\% Conference Abstract/Paper (9,931 of
10,094).
All other publishers are dominated by Research Articles, ranging from 68\% (Wiley) to 97\% (IOS Press).
Wiley has the highest Review Article (17\%) and Meta-Analysis (7\%) share.
}%

\begin{figure*}[!htbp]
	\centering
	\includegraphics[width=\linewidth]{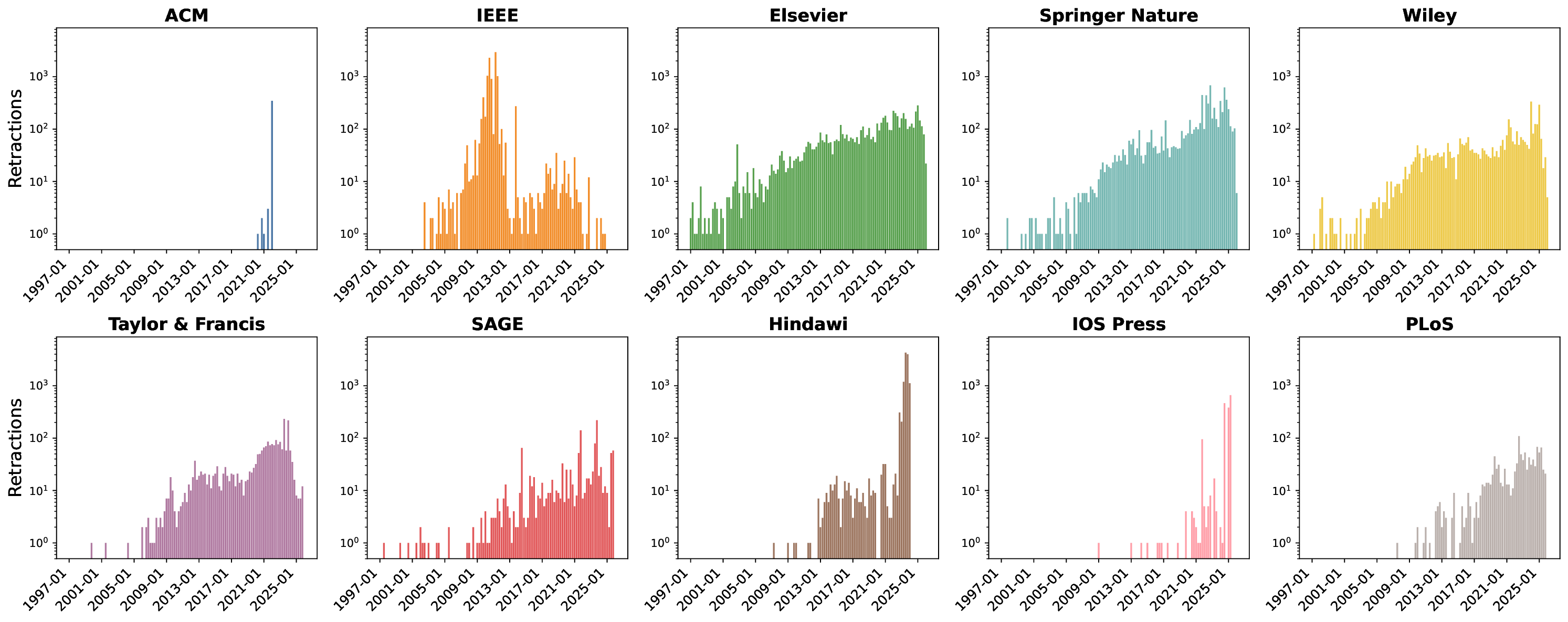}
	\caption{Number of retractions per quarter (log scale) for each publisher. The logarithmic y-axis highlights differences in magnitude across publishers.}
	\label{fig:entries-per-quarter-log}
\end{figure*}

\subsection{Temporal Trends}%
\label{sec:temporal-trends}%

\notouch{%
Figure~\ref{fig:entries-per-quarter-log} displays quarterly retraction counts for each publisher on a logarithmic scale.
IEEE retraction activity spans 2002--2026, with the highest quarterly counts in 2022--2023.
ACM's timeline begins later: its first entry in the Retraction Watch database is dated 2020.
The bulk of ACM's retractions (347 of 354) are concentrated in Q3 2021, when two conference proceedings were retracted~\cite{more-than-300-at-once}, with Compromised Peer Review as the dominant reason.
Outside of this event, ACM's quarterly counts are negligible.
}%

\notouch{%
Elsevier and Springer Nature show steady growth over the past decade.
Hindawi displays a sharp spike in 2022--2023, consistent with the mass retractions following Wiley's acquisition~\cite{d41586-023-03974-8.pdf}.
IOS Press and PLoS show retraction activity spanning the early 2010s onward.
SAGE exhibits a surge around 2015, coinciding with well-documented peer review manipulation events~\cite{ferguson2014peer-review-scam}.
}%

\subsection{Retraction Reasons}%
\label{sec:retraction-reasons}%
\notouch{%
Using the consolidated reason categories from Section~\ref{sec:reason-consolidation}, we examined retraction reason distributions across publishers.
Meta-reasons (e.g., ``Investigation by Journal/Publisher,'' ``Information'') were excluded to focus on substantive causes.
}%

\begin{figure*}[!htbp]
	\centering
	\includegraphics[width=\linewidth]{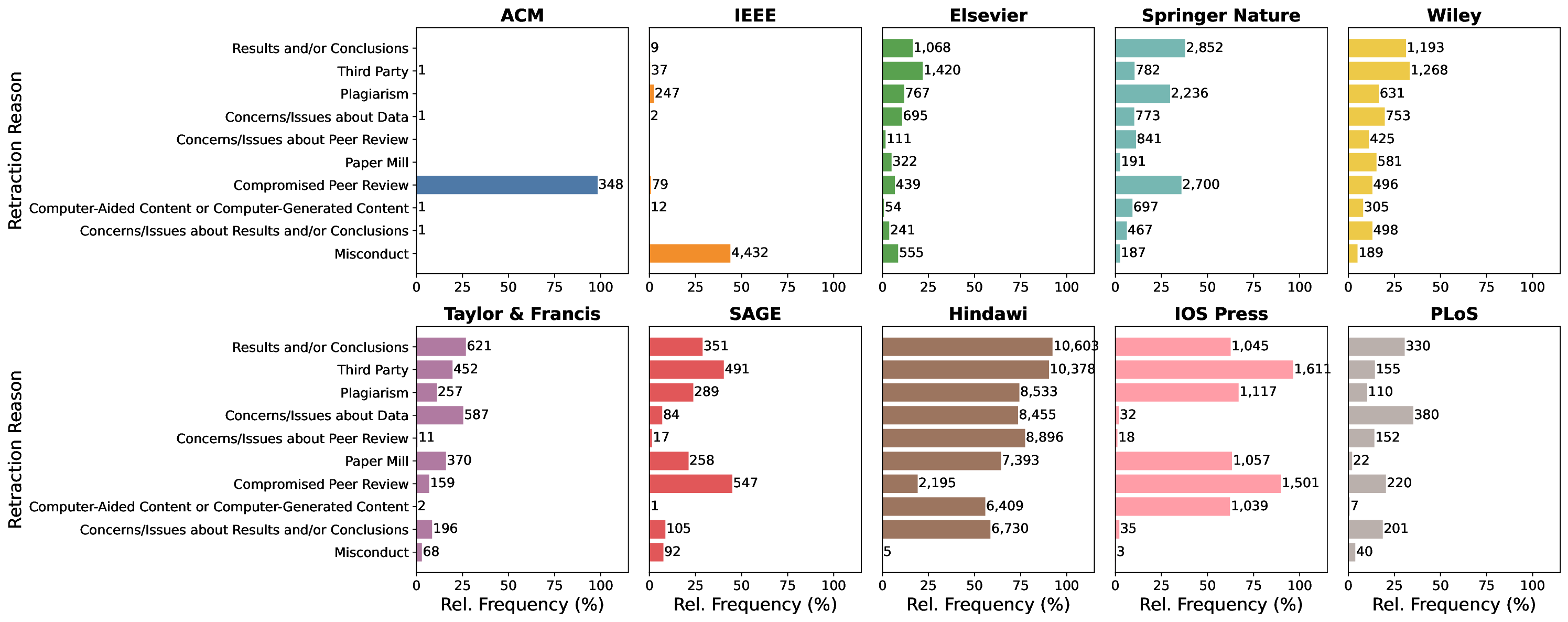}
	\caption{Top ten consolidated retraction reasons per publisher. Bars show relative frequency (\%). Absolute counts are annotated at bar ends. Meta-reasons (e.g., Investigation by Journal/Publisher) are excluded.}
	\label{fig:reasons-by-publisher}
\end{figure*}

\notouch{%
Across all publishers, the most prevalent reasons are Results and/or Conclusions (39.2\%; 18,072 retractions), Third Party (36.0\%; 16,595), Plagiarism (30.8\%; 14,187), Data Concerns (25.5\%; 11,762), and Peer Review Concerns (22.7\%; 10,471).
A retraction may list multiple reasons, so percentages sum to more than 100\%.
Table~\ref{tab:reasons-by-publisher-top3} and Figure~\ref{fig:reasons-by-publisher} show that per-publisher profiles differ.
}%

\notouch{%
ACM is dominated by a single reason: 98.3\% of retractions (348 of 354) cite Compromised Peer Review, the vast majority from the Q3 2021 event described above.
IEEE's leading reason is Misconduct (44\%; 4,432), followed by Removed (12\%) and Plagiarism (2\%).
Springer Nature's profile is shaped by Results and/or Conclusions (38\%), Compromised Peer Review (36\%), and Plagiarism (30\%).
SAGE shows Compromised Peer Review (45\%) alongside Third Party involvement (40\%).
Hindawi's profile reflects its mass retraction events: Results and/or Conclusions (92\%), Third Party (90\%), and Peer Review Concerns (77\%) are each cited in the majority of retractions.
IOS Press shows a similarly concentrated pattern, with Third Party (96\%) and Compromised Peer Review (90\%) dominating.
PLoS displays a more distributed profile, led by Data Concerns (35\%), Image Duplication (33\%), and Results and/or Conclusions (31\%).
}%

%

\newcommand{\acmHHI}{0.94}
\newcommand{\acmEntropy}{0.21}
\newcommand{\acmPermP}{< 0.001}
\newcommand{\acmChiSq}{5858.1}
\newcommand{\acmChiDof}{10}
\newcommand{\acmChiP}{< 0.001}
\newcommand{\acmChiPAsymptotic}{< 0.001}

\begin{table}[htbp]
  \centering
  \caption{Reason concentration per publisher. HHI (Herfindahl-Hirschman Index) ranges from $1/K$ (uniform) to 1.0 (single category). Shannon entropy $H$ is higher for more diverse profiles. Permutation $p$-values test whether the observed HHI could arise from randomly drawing $N$ entries from the pooled distribution ($B = 10{,}000$).}
  \label{tab:concentration}
  \small
  \begin{tabular}{l r r r r}
    \toprule
    Publisher & $N$ & HHI & $H$ & Perm.\ $p$ \\
    \midrule
    ACM & 354 & 0.9397 & 0.21 & < 0.001 \\
    IEEE & 10,094 & 0.5434 & 0.99 & < 0.001 \\
    IOS Press & 1,671 & 0.1427 & 2.13 & < 0.001 \\
    Hindawi & 11,486 & 0.1090 & 2.34 & < 0.001 \\
    SAGE & 1,217 & 0.0763 & 3.08 & < 0.001 \\
    Springer Nature & 7,534 & 0.0573 & 3.27 & 1.000 \\
    Taylor \& Francis & 2,314 & 0.0544 & 3.27 & 1.000 \\
    Wiley & 3,817 & 0.0479 & 3.42 & 1.000 \\
    PLoS & 1,080 & 0.0444 & 3.38 & 1.000 \\
    Elsevier & 6,520 & 0.0361 & 3.62 & 1.000 \\
    \bottomrule
  \end{tabular}
\end{table}

%

\notouch{%
To quantify these differences, we computed the Herfindahl-Hirschman Index (HHI) and Shannon entropy for each publisher's reason-mention shares (see Table~\ref{tab:concentration}).
ACM's HHI (\acmHHI) is the highest in the dataset, and its Shannon entropy (\acmEntropy) is the lowest, confirming that its reason profile is the most concentrated among all ten publishers.
IEEE also shows elevated concentration (HHI = 0.54) due to the dominance of Misconduct, followed by IOS Press (0.14) and Hindawi (0.11). The remaining publishers fall below an HHI of 0.08.
A permutation test ($B = 10{,}000$) assessed whether each publisher's HHI could arise from randomly drawing $N$ retractions from the pooled distribution of all 46,087 retractions.
ACM's observed concentration falls outside the null distribution ($p < 0.001$), ruling out small sample size as an explanation. IEEE, Hindawi, IOS Press, and SAGE also show significant concentration ($p < 0.001$), but none approach ACM's level.
A chi-square goodness-of-fit test further confirms that ACM's reason distribution differs from the pooled remaining publishers ($\chi^2 = \acmChiSq$, $\text{df} = \acmChiDof$, Monte Carlo $p\ \acmChiP$).
}%

\begin{table*}[!htbp]
	\centering
	\caption{Top three consolidated retraction reasons per publisher (percentage of retractions within that publisher). Meta-reasons excluded. Reason names are abbreviated for space.}
	\label{tab:reasons-by-publisher-top3}
	\small
	\begin{tabular}{l r r l r l r l}
		\toprule
		Publisher & N & \multicolumn{2}{l}{1st reason} & \multicolumn{2}{l}{2nd reason} & \multicolumn{2}{l}{3rd reason} \\
		\midrule
		ACM & 354 & 98\% & Compr. Peer Rev. & 0\% & Data Concerns & 0\% & Results Concerns \\
		IEEE & 10,094 & 44\% & Misconduct & 12\% & Removed & 2\% & Plagiarism \\
		Elsevier & 6,520 & 22\% & Third Party & 16\% & Results/Concl. & 13\% & Removed \\
		Springer Nature & 7,534 & 38\% & Results/Concl. & 36\% & Compr. Peer Rev. & 30\% & Plagiarism \\
		Wiley & 3,817 & 33\% & Third Party & 31\% & Results/Concl. & 20\% & Data Concerns \\
		Taylor \& Francis & 2,314 & 27\% & Results/Concl. & 25\% & Data Concerns & 20\% & Third Party \\
		SAGE & 1,217 & 45\% & Compr. Peer Rev. & 40\% & Third Party & 29\% & Results/Concl. \\
		Hindawi & 11,486 & 92\% & Results/Concl. & 90\% & Third Party & 77\% & Peer Rev. Concerns \\
		IOS Press & 1,671 & 96\% & Third Party & 90\% & Compr. Peer Rev. & 67\% & Plagiarism \\
		PLoS & 1,080 & 35\% & Data Concerns & 33\% & Image Dupl. & 31\% & Results/Concl. \\
		\bottomrule
	\end{tabular}
\end{table*}

\begin{figure*}[!htbp]
  \centering
  \includegraphics[width=\textwidth]{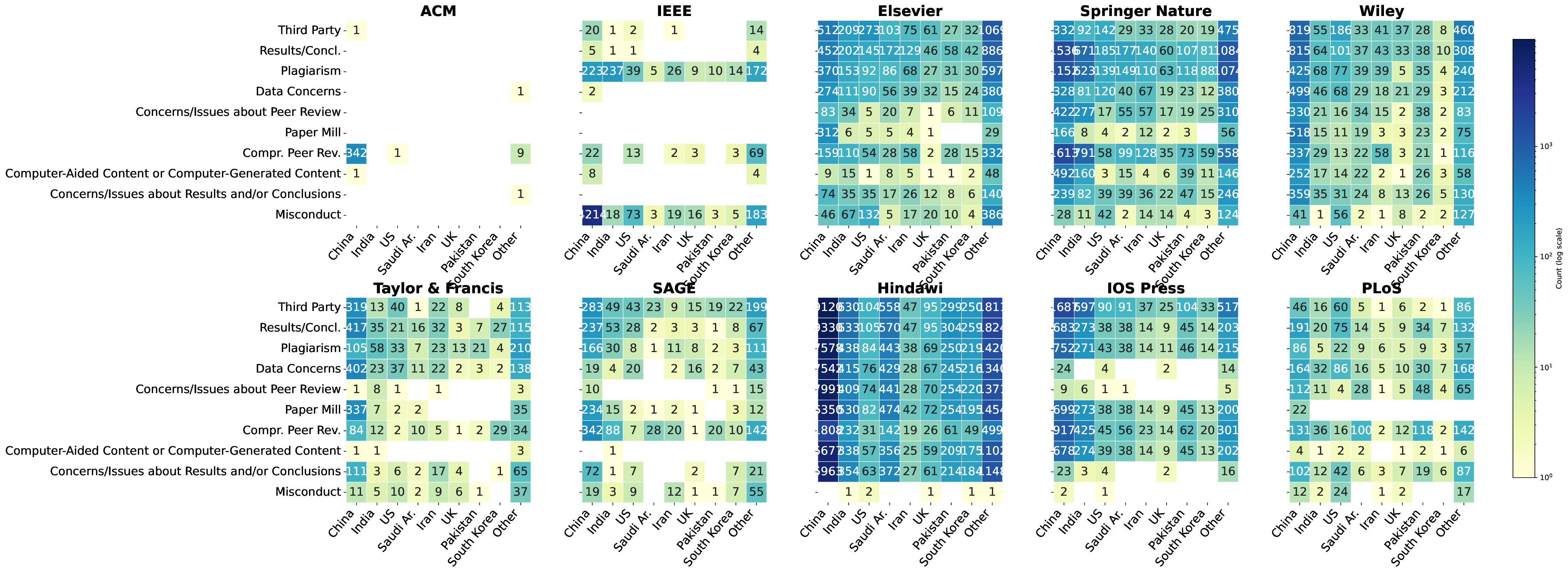}
  \caption{Top ten retraction reasons by top ten author countries for each publisher. Cell values indicate co-occurrence counts; color intensity follows a logarithmic scale. Empty cells denote zero co-occurrences.}
  \label{fig:heatmap-panel}
\end{figure*}

\notouch{%
The per-publisher correlation plots in Figure~\ref{fig:heatmap-panel} cross-tabulate retraction reasons with author countries.
ACM's heatmap is dominated by a single cell: China $\times$ Compromised Peer Review (342 retractions), reflecting the Q3 2021 bulk retraction.
No ACM retractions appear for Misconduct, Plagiarism, Results and/or Conclusions, Rogue Editor, Image Duplication, or Author Issues. These categories feature prominently for other publishers.
IEEE shows a China-dominated but reason-diverse pattern: the largest cell is China $\times$ Misconduct (4,213), followed by China $\times$ Removed (1,135) and India $\times$ Plagiarism (237).
SAGE is distinctive in that the United States features prominently alongside China, with Third Party and Data Concerns disproportionately represented.
Springer Nature displays a Rogue Editor cluster from India that is largely unique among the publishers examined.
}%

\subsection{Geographic Distribution}%
\label{sec:geographic-distribution}%

\begin{figure*}[!htbp]
	\centering
	\includegraphics[width=\linewidth]{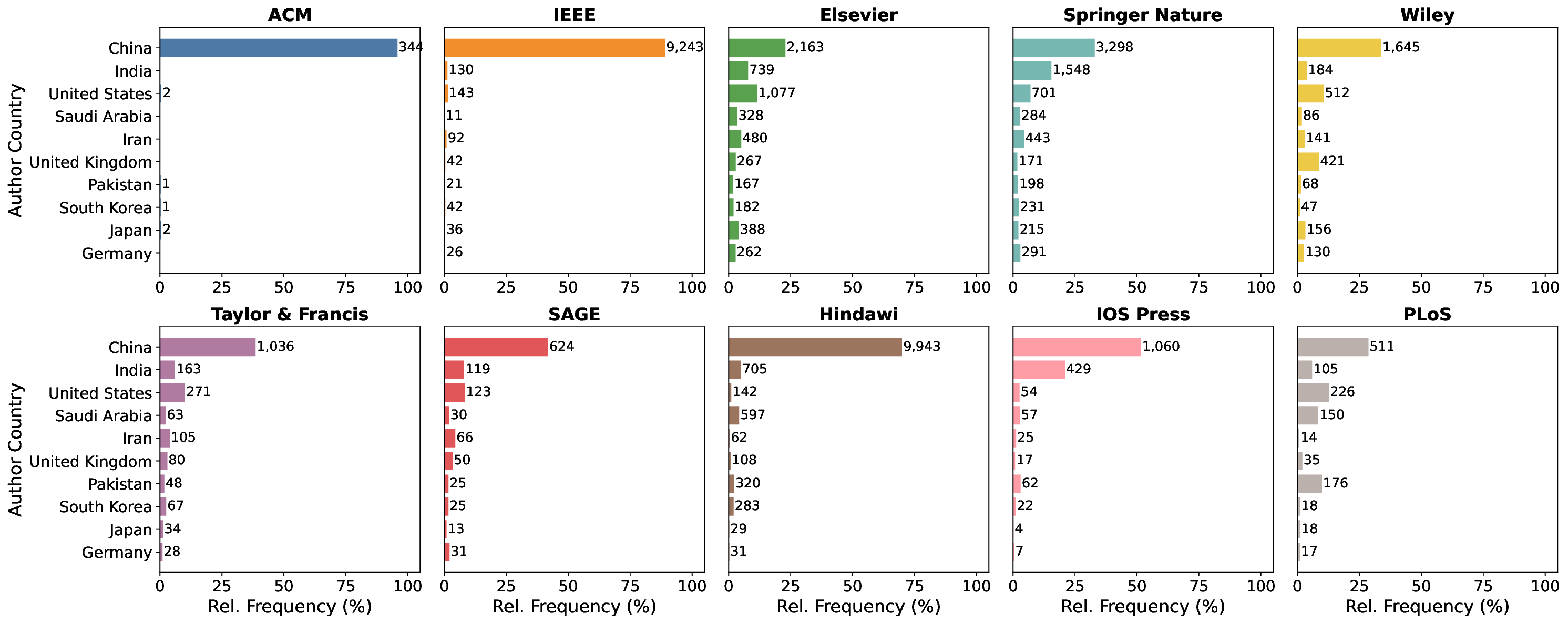}
	\caption{Top 10 author countries per publisher (relative frequency, \%). Country affiliations are expanded; retractions listing multiple countries contribute to each listed country.}
	\label{fig:author-countries}
\end{figure*}

\notouch{%
Author country affiliations were expanded so that retractions listing multiple countries are counted once per country, yielding 56,847 affiliations across 172 unique countries (excluding ``Unknown'').
Table~\ref{tab:top-countries} lists the top 10 author country affiliations.
China dominates with 29,867 affiliations (52.54\%), followed by India (4,122; 7.25\%) and the United States (3,251; 5.72\%).
Saudi Arabia (2.83\%), Iran (2.51\%), and the United Kingdom (2.10\%) round out the top six.
For comparison, China-affiliated research organizations account for approximately 16.5\% of the combined publication output of these ten publishers in Dimensions (1997--2026), placing China's retraction share at roughly four times its publication base rate (Table~\ref{tab:china-base-rate}).
China is the leading country at every publisher, though the degree of dominance varies (see Figure~\ref{fig:author-countries}).
}%

\begin{table}[!htbp]
	\caption{Top 10 author countries by retraction affiliation count (expanded; retractions with multiple country affiliations are counted once per country); Domestic vs.\ international authorship breakdown for the top 10 retraction countries.}
	\label{tab:top-countries}
	\label{tab:solo-multi-country}
	\centering
	\small
	\begin{tabular}{lrrcrrr}
		\toprule
		&       &            & ~ & \multicolumn{3}{c}{Authorship} \\
		Country & Count & Percentage & ~ & Domestic & International & Domestic (\%) \\
		\midrule
		China          & 29,867 & 52.54\% & ~ & 27,274 & 2,593 & 91.3 \\
		India          &  4,122 &  7.25\% & ~ & 2,563 & 1,559 & 62.2 \\
		United States  &  3,251 &  5.72\% & ~ & 1,681 & 1,570 & 51.7 \\
		Saudi Arabia   &  1,606 &  2.83\% & ~ & 227 & 1,379 & 14.1 \\
		Iran           &  1,428 &  2.51\% & ~ & 920 & 508 & 64.4 \\
		United Kingdom &  1,191 &  2.10\% & ~ & 477 & 714 & 40.1 \\
		Pakistan       &  1,086 &  1.91\% & ~ & 178 & 908 & 16.4 \\
		South Korea    &    918 &  1.61\% & ~ & 425 & 493 & 46.3 \\
		Japan          &    895 &  1.57\% & ~ & 658 & 237 & 73.5 \\
		Germany        &    823 &  1.45\% & ~ & 446 & 377 & 54.2 \\
		\bottomrule
	\end{tabular}
\end{table}

%

\begin{table}[!htbp]
\caption{China-affiliated publication share vs.\ retraction share per publisher.
Publication counts from Dimensions (1997--2026); retraction counts from Retraction Watch.
The ratio divides the retraction share by the publication share; values above 1.0
indicate overrepresentation in retractions relative to publication output.}
\label{tab:china-base-rate}
\centering
\small
\begin{tabular}{lrrr}
\toprule
Publisher & China Pub \% & China Ret \% & Ratio \\
\midrule
IOS Press            & 0.1\%  & 63.4\% & 940.04 \\
SAGE                 & 7.6\%  & 51.3\% & 6.78 \\
ACM                  & 16.9\% & 97.2\% & 5.74 \\
Taylor \& Francis     & 9.6\%  & 44.8\% & 4.65 \\
IEEE                 & 25.9\% & 91.6\% & 3.54 \\
Wiley                & 12.5\% & 43.1\% & 3.44 \\
PLoS                 & 14.1\% & 47.3\% & 3.37 \\
Hindawi              & 28.2\% & 86.6\% & 3.07 \\
Springer Nature      & 15.5\% & 43.8\% & 2.82 \\
Elsevier             & 17.8\% & 33.2\% & 1.87 \\
\midrule
\textbf{Aggregate} & 16.5\% & 64.8\% & 3.92 \\
\bottomrule
\end{tabular}
\end{table}

\subsection{Country Co-Authorship}%
\label{sec:coauthorship}%

\begin{figure}[!htbp]
	\centering
	\includegraphics[width=.7\linewidth]{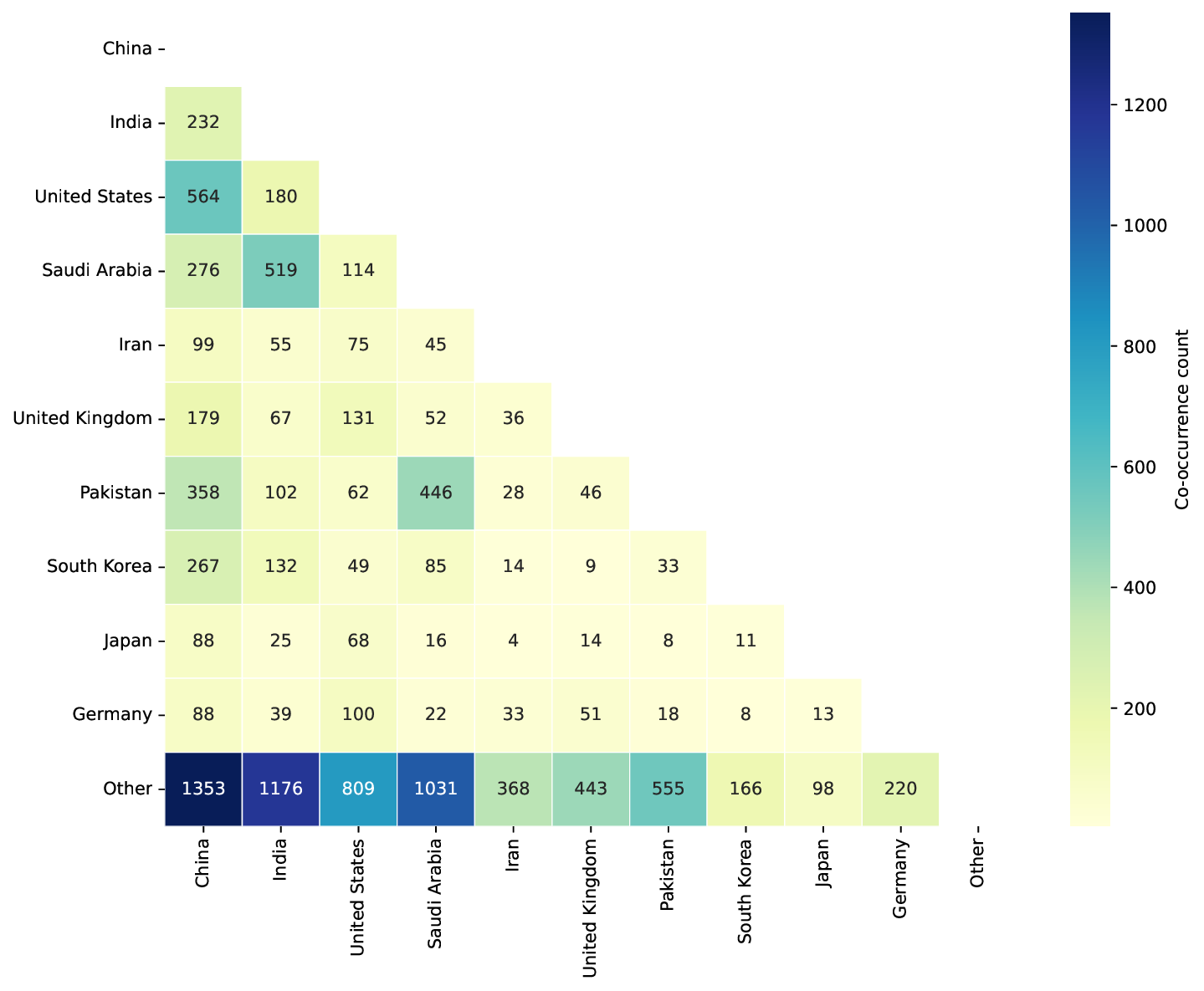}
	\caption{Country co-occurrence in retracted publications (lower triangle). Each cell shows the number of retractions listing both countries. Remaining countries are aggregated under ``Other.''}
	\label{fig:country-cooccurrence-heatmap}
\end{figure}

\notouch{%
Of all 46,087 retractions, 39,087 (84.8\%) list authors from a single country (domestic authorship) and 7,000 (15.2\%) list authors from multiple countries (international authorship).
Table~\ref{tab:solo-multi-country} shows the breakdown for the top 10 countries.
China has a 91.3\% domestic authorship rate, while the United States is substantially lower at 51.7\%.
Saudi Arabia is an outlier at only 14.1\% domestic retractions, reflecting heavy international co-authorship. Pakistan shows a similarly low rate (16.4\%).
}%

\notouch{%
Of the 3,251 retractions involving US-affiliated authors, 1,681 (51.7\%) are solo-US, 564 (17.3\%) are US--China co-authorships, and 1,006 (30.9\%) are US co-authorships with countries other than China.
In US--China co-authored retractions, China is listed as the first country in 93.8\% of cases (529 of 564).
When country rankings are computed using only the first-listed country rather than expanded affiliations, the absolute US count drops from 3,251 to 1,697 (a 48\% reduction), though the rank (third, behind China and India) remains stable.
}%

\notouch{%
The highest co-occurrence counts involve China paired with other countries, followed by US--China (564) and Saudi Arabia--India (Figure~\ref{fig:country-cooccurrence-heatmap}).
}%

\begin{figure}[!htbp]
	\centering
	\includegraphics[width=\linewidth]{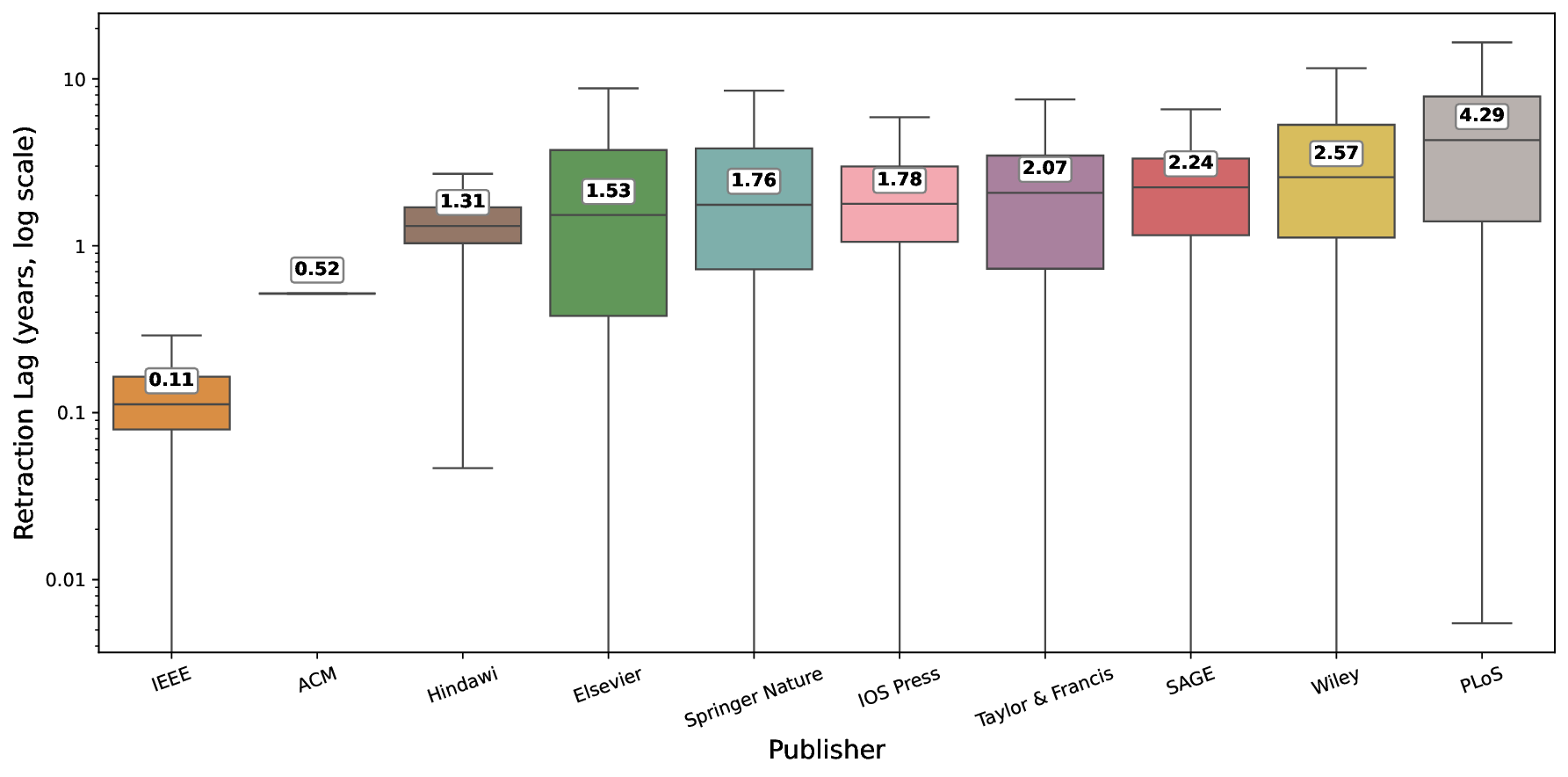}
	\caption{Retraction lag (in years, log scale) by publisher. Boxes show interquartile range; annotated values indicate the median. Outliers are suppressed for readability. The full statistics are in Appendix \ref{appendix:lag}.}
	\label{fig:retraction-lag-boxplot}
\end{figure}

\subsection{Retraction Lag}%
\label{sec:retraction-lag}%
\notouch{%
Retraction lag (the time between original publication and retraction) varies across publishers (see Figure~\ref{fig:retraction-lag-boxplot} and Table~\ref{tab:retraction-lag} in Appendix~\ref{appendix:lag}).
IEEE has the shortest median lag at 41 days, with 99.2\% of retractions occurring within one year.
ACM's median of 189 days is 4.6 times longer, though 98.3\% of ACM retractions still fall within the one-year window.
The remaining publishers have longer retraction lags: Hindawi (median 480 days), Elsevier (558), Springer Nature (643), IOS Press (651), Taylor~\&~Francis (758), SAGE (818), Wiley (940), and PLoS (1,568 days, i.e., 4.3 years).
PLoS has the highest mean lag (1,825 days; 5.0 years).
SAGE has the greatest variability (SD = 1,496 days), reflecting a broad range of retraction timelines.
}%

\subsection{Repeat Authors}%
\label{sec:repeat-authors}%
\notouch{%
Using Dimensions \texttt{researcher\_id} and name+institution disambiguation (see Section~\ref{sec:normalization}), we identified 148,508 unique author identifiers (see Table~\ref{tab:repeat-authors-per-publisher}).
The majority (132,900; 89.49\%) appear in exactly one retraction.
A total of 15,608 (10.51\%) have two or more retractions, 1,482 (1.00\%) have five or more, and 13 (0.01\%) have 50 or more.
Additionally, 5,661 authors appear across two or more publishers, of whom 855 have five or more total retractions.
This overlap suggests that integrity concerns are not confined to individual publishers.
}%

\begin{table}[!htbp]
	\centering
	\caption{Per-publisher repeat-author statistics. Columns show the number of unique authors, the count and percentage with 2+ and 5+ retractions within that publisher, and the mean retractions per author.}
	\label{tab:repeat-authors-per-publisher}
	\small
	\begin{tabular}{l r r r r r r}
		\toprule
		Publisher & Authors & 2+ & 2+ (\%) & 5+ & 5+ (\%) & Mean \\
		\midrule
		Hindawi & 37,781 & 2,712 & 7.2 & 164 & 0.4 & 1.11 \\
		Springer Nature & 28,277 & 2,403 & 8.5 & 210 & 0.7 & 1.17 \\
		Elsevier & 26,263 & 2,388 & 9.1 & 321 & 1.2 & 1.21 \\
		IEEE & 24,112 & 1,868 & 7.7 & 56 & 0.2 & 1.11 \\
		Wiley & 14,965 & 1,712 & 11.4 & 162 & 1.1 & 1.22 \\
		Taylor \& Francis & 8,224 & 458 & 5.6 & 35 & 0.4 & 1.10 \\
		PLoS & 6,321 & 617 & 9.8 & 67 & 1.1 & 1.19 \\
		IOS Press & 4,715 & 372 & 7.9 & 46 & 1.0 & 1.16 \\
		SAGE & 3,888 & 223 & 5.7 & 28 & 0.7 & 1.12 \\
		ACM & 626 & 32 & 5.1 & 0 & 0.0 & 1.06 \\
		\bottomrule
	\end{tabular}
\end{table}

\notouch{%
ACM has the lowest repeat-offender rate: 5.1\% of the 626 authors on its retracted papers have two or more retractions, and none have five or more.
Wiley has the highest rate, with 11.4\% of its 14,965 authors having two or more retractions.
IEEE, despite the fourth-largest author pool (24,112), has a low repeat rate (7.7\% with 2+) and one of the lowest rates of five or more (0.2\%).
}%

\notouch{%
The most-retracted individual in the dataset is Joachim Boldt with 129 retractions across three publishers (Elsevier, Springer Nature, and Wiley).
Other authors with more than 90 retractions include Hironobu Ueshima (112; three publishers), Yoshitaka Fujii (98; four publishers), and Ali Nazari (93; four publishers).
All four appear exclusively at Elsevier, Springer Nature, and Wiley.
}%


\section{Discussion}%
\label{sec:discussion}%
\notouch{%
Retraction practices differ across the ten publishers not only in volume but in composition, timing, and reason diversity. These differences reflect distinct editorial cultures, detection strategies, and policy choices.
We first discuss the patterns that emerge across all ten publishers (Section~\ref{sec:disc:landscape}), then examine ACM's retraction record (Section \ref{sec:acm-record}) as a case that raises specific questions about threshold-setting (Section~\ref{sec:disc:threshold}), transparency (Section~\ref{sec:disc:transparency}), and implications for retraction practice (Section~\ref{sec:disc:implications}).
}%

\subsection{Retraction Patterns in a Multi-Publisher Landscape}%
\label{sec:disc:landscape}%

\subsubsection{Comparison of publishers}

\notouch{%
The 10-publisher comparison reveals variation in how retraction policies are practiced.
Publishers differ in volume, reasons cited, retraction lag, geographic distribution, and repeat-offender rates.
Each publisher's record reflects its editorial practices, disciplinary composition, and policy choices.
Several publisher-specific patterns deserve attention.
}%

\paragraph{Hindawi.} \notouch{%
Hindawi's record illustrates what happens when a publisher discovers systemic compromise and acts on it. The reason profile and temporal surge (Section~\ref{sec:retraction-reasons}, Section~\ref{sec:temporal-trends}) are consistent with mass retraction events targeting paper mill output, triggered by Wiley's acquisition and subsequent investigations~\cite{d41586-023-03974-8.pdf}. The result is a retraction record shaped
by episodic discovery.%
}%

\paragraph{IEEE.} \notouch{%
IEEE's proactive approach, publicly announced in 2019~\cite{retractions-from-xplore.html}, demonstrates that large-scale retraction is operationally feasible. The shortest median lag and highest within-one-year completion rate in the dataset (Section~\ref{sec:retraction-lag}) confirm that retraction at scale need not be protracted. IEEE's reason profile is also among the most diverse, with Misconduct, Removed, and Plagiarism all represented.%
}%

\paragraph{SAGE.} \notouch{%
SAGE's record suggests a different adjudication culture, one that uses expressions of concern as an intermediate step while investigations proceed. The pre-filtering breakdown (Table~\ref{tab:retraction-nature}) shows that SAGE's expression-of-concern rate (27.3\%) is the highest in the dataset, which may reflect a lower threshold for flagging questionable work even when the evidence is not yet sufficient for full retraction. The reason profile reflects the well-documented peer review manipulation events of 2014~\cite{ferguson2014peer-review-scam}.
}%

\paragraph{PLoS.} \notouch{%
PLoS combines the longest median retraction lag in the dataset with the most distributed reason profile (Section~\ref{sec:retraction-reasons}, Section~\ref{sec:retraction-lag}). This combination suggests a publisher that encounters a broad range of integrity issues but resolves them over extended timelines. The pre-filtering breakdown (Table~\ref{tab:retraction-nature}) also shows PLoS with the second-highest expression-of-concern rate (15.4\%) and the highest correction share (6.2\%), consistent with a publisher that uses the full range of corrective actions.
}%

\paragraph{Elsevier.} \notouch{%
Despite the largest publication volume, Elsevier has the lowest normalized retraction rate. What distinguishes Elsevier is its reinstatement record: the pre-filtering breakdown (Table~\ref{tab:retraction-nature}) shows that the large majority of reinstatements in the dataset originate from Elsevier (86 of 98, or 87.8\%). A high reinstatement rate indicates that some retraction decisions are reversed upon further review, which may reflect both a willingness to retract and a willingness to correct course when the evidence warrants it.
}%

\paragraph{IOS Press.} \notouch{%
IOS Press shows a pattern consistent with organized manipulation of the peer review process, similar to Hindawi's but at a different scale. Its high normalized rate (Section~\ref{sec:dataset-overview}) reflects mass retractions against a small publication base~\cite{kincaid2024iospress}, with a reason profile dominated by Third Party involvement and Compromised Peer Review.
}%

\subsubsection{Geographic patterns and retraction lag}

\notouch{%
China-affiliated authors account for the largest share of retractions at every publisher, a systemic phenomenon not attributable to any single venue. This overrepresentation persists after normalizing for publication volume: China accounts for roughly one-sixth of publications at these ten publishers but half of retraction affiliations (Table~\ref{tab:china-base-rate}). This pattern aligns with prior research on incentive structures~\cite{zhang-wang-2024-research-misconduct-in-china-towards-an-institutional-analysis.pdf, Elite} and paper mills~\cite{else2021.pdf}. China's 91.3\% domestic authorship rate indicates that the retractions predominantly involve domestic research teams rather than international collaborations. Saudi Arabia (14.1\% domestic authorship) and Pakistan (16.4\%) show the opposite pattern, with the large majority of retracted work involving international co-authorship.
}%

\notouch{%
Median retraction lag spans a 40-fold range across the ten publishers. 
These differences likely reflect a combination of editorial capacity, investigation complexity, and policy choices about when to act on incomplete evidence.
}%

\subsection{ACM's retraction record}%
\label{sec:acm-record}%

\notouch{%
Within this cross-publisher landscape, ACM occupies a distinctive position.
The normalized retraction rate offers a necessary corrective to the raw count disparity (e.g., 354 ACM  vs.\ 10,094 IEEE).
Scaled by publication volume, ACM's rate of 5.65 retractions per 10,000 publications places it eighth of ten, comparable to SAGE (5.46) and above only Elsevier (3.97).
ACM's retraction record is not negligible in proportion to its output.
}%

\notouch{%
What distinguishes ACM from every other publisher is not the rate but the \emph{composition}.
ACM's reason profile is the most concentrated in the dataset by both HHI and Shannon entropy (Table~\ref{tab:concentration}), and a permutation test confirms that this concentration cannot be explained by ACM's smaller sample size ($p < 0.001$; Section~\ref{sec:retraction-reasons}).
Even publishers with a clearly dominant category show secondary and tertiary reasons that collectively account for a substantial share.
ACM's record, by contrast, is dominated by a single reason stemming predominantly from one single event~\cite{more-than-300-at-once}.
The ten publishers in this dataset span different organizational types and disciplinary scopes, from single-discipline society publishers to multi-disciplinary commercial houses, and these structural differences complicate direct comparison of reason profiles.
The comparison most resistant to this confound is IEEE, which shares ACM's disciplinary focus, conference-heavy publication model, and professional society structure, yet shows one of the most diverse reason profiles in the dataset.
These compositional findings do not depend on coverage completeness. Even if Retraction Watch under-counts ACM retractions, the reason distribution among the 354 recorded entries is independently informative: unrecorded retractions would need to span the reason categories absent from ACM's profile to alter the pattern.
}%

\notouch{%
The temporal dimension reinforces ACM's outlier status among the investigated publishers.
ACM's first retraction dates to 2020, despite the Digital Library launching in 1997~\cite{ACMDL-launch}. The vast majority of retractions cluster in Q3 2021.
In contrast, Elsevier, Springer Nature, Wiley, and IEEE have retraction notices dating back to the early 2000s,
and IOS Press and PLoS have retraction notices dating back to the early 2010s.
}%

\notouch{%
The systemic pressures that drive misconduct elsewhere are not publisher-specific; they affect the scholarly ecosystem as a whole.
The question is not whether such pressures exist within ACM's author community (they almost certainly do), 
but whether ACM's retraction record reflects them.
ACM's retraction record is small enough that a single event dominates the distribution. This is not a sampling limitation; it is the phenomenon under study. The question is not whether the record is thin, but what a record this thin implies for a publisher of ACM's scale and longevity.
}%

\subsection{ACM's Retraction Threshold}%
\label{sec:disc:threshold}%

\notouch{%
ACM's policy describes its retraction threshold as ``extremely high''~\cite{ACM-POLICY}.
The data invite reflection on what a record under that threshold looks like.
The interpretations that follow are abductive: we identify the explanations most consistent with the observed patterns, while acknowledging that bibliometric data alone cannot adjudicate among them.
Three explanations for ACM's outlier profile merit consideration.
}%

\paragraph{ACM genuinely has less misconduct.}
\notouch{%
It is possible that ACM-published research contains fewer of the problems that lead to retraction elsewhere.
If so, the sparse record would be a positive signal.
But the data make this difficult to sustain at the observed scale.
ACM publishes across the same broad fields (computer science, information systems, human--computer interaction) and draws from an overlapping author population as other publishers in this dataset.
The systemic pressures that drive misconduct, including publish-or-perish incentives~\cite{edwards-roy-2017-academic-research-in-the-21st-century-maintaining-scientific-integrity-in-a-climate-of-perverse.pdf, tijdink2014.pdf} and paper mills~\cite{else2021.pdf}, are not confined to specific publishers.
Seven of the top ten global retraction reasons are absent from ACM's record (Table~\ref{tab:reasons-by-publisher-top3}), despite affecting publishers of comparable size and scope.
A chi-square goodness-of-fit test comparing ACM's reason distribution to the pooled remaining publishers confirms this divergence ($\chi^2 = \acmChiSq$, $\text{df} = \acmChiDof$, $p\ \acmChiP$).
If the systemic pressures that drive retractions at other publishers are not publisher-specific, the expected baseline would include at least some retractions in these categories. Their complete absence is a pattern that requires explanation.
This is not an inference from absence to presence. The data do not establish that plagiarism or fabrication occurs in ACM-published work. They establish that ACM's retraction profile diverges compositionally from every other publisher in the dataset, a pattern that warrants examination regardless of the underlying cause.
}%

\paragraph{Detection mechanisms differ.}
\notouch{%
A second possibility is that misconduct occurs in ACM-published work but goes undetected at higher rates.
ACM's record is consistent with a reactive, case-by-case approach: the 2021 bulk retraction was triggered by a large-scale peer review compromise~\cite{more-than-300-at-once} that was presumably difficult to overlook.
IEEE, by contrast, publicly announced a proactive approach in 2019~\cite{retractions-from-xplore.html}, and its retraction activity surged in the years that followed.
Proactive investigation surfaces misconduct that a purely reactive approach would miss.
Detection differences alone, however, cannot fully account for the \emph{reason-type} gap.
Even under a slower, reactive regime, one would expect at least some retractions for plagiarism, data fabrication, or unreliable results over decades of publishing.
That ACM's record contains essentially none suggests detection is not the full explanation.%
}%

\paragraph{Policy as a shaping factor.}
\notouch{%
The most parsimonious reading of the data is that ACM's threshold excludes the vast majority of cases that other publishers would retract.
The 2021 conference proceeding retraction cleared this threshold because it involved a large-scale, unambiguous compromise, a category of misconduct that is straightforward to establish.
Individual cases of plagiarism, data fabrication, or flawed results require more nuanced adjudication and are more easily contested.
Under a policy that demands ``extremely high'' certainty, such cases are less likely to result in formal retraction.
}%

\notouch{%
A publisher's retraction record reflects not only misconduct prevalence but also willingness and capacity to act.
The COPE retraction guidelines, which ACM has adopted for a subset of its journals, recommend retraction when there is ``clear evidence that the findings are unreliable'' or evidence of misconduct~\cite{retraction-guidelines-cope.pdf}, a substantially lower threshold than ``extremely high.''
When a publisher sets a high bar, the expected pattern is fewer retractions and a record that may under-represent the true extent of problems.
This is not a neutral outcome.
Retracted articles continue to be cited long after retraction~\cite{qss_a_00303.pdf, s11192-022-04321-w.pdf, qss_a_00222.pdf}, retracted findings propagate into public policy~\cite{qss_a_00243.pdf} and social media~\cite{3637335.pdf}, and the financial costs of flawed research are substantial~\cite{elife-02956-v1.pdf}.
When problematic work is not retracted, these harms compound: no formal notice is issued, and the scholarly community has no mechanism to identify and disregard the flawed findings.
}%

\subsection{Transparency and the Dark Archive}%
\label{sec:disc:transparency}%

\notouch{%
Separate from the retraction threshold is ACM's ``dark archive'' for removed works~\cite{ACM-POLICY}.
When ACM removes a publication, it moves to a private archive.
The scholarly community cannot verify which works have been removed, the reasons for removal, or the scope of the corrections applied.
}%

\notouch{%
A retraction notice informs readers that a work should not be relied upon, allowing downstream researchers to reassess findings built on it.
When works are removed without public notice, the corrective function is served for the publisher but not for the broader community.
Researchers who previously cited a removed work may have no way of knowing it has been withdrawn.
}%

\notouch{%
The paywall data offer a useful contrast.
ACM has 0\% paywalled retractions.
All retraction-related content identified by Retraction Watch is openly accessible.
This is a positive finding.
But the dark archive introduces a different form of opacity: openness with respect to \emph{known} retracted works coexists with opacity regarding works removed from public view entirely.
The net effect on research integrity depends on volumes the community cannot assess, because the archive's contents are, by design, unknown.
}%

\notouch{%
The scholarly community's recent investments in transparency
 	reinforce this concern.
Crossref's acquisition of the Retraction Watch database~\cite{Crossref} reflects a consensus that retraction metadata should be openly available for indexing, citation analysis, and integrity monitoring~\cite{retractionwatch}.
A dark archive moves in the opposite direction.
ACM making its metadata publicly available (titles, DOIs, dates, and reasons for removal) would allow the community to assess the scope of corrections in the ACM Digital Library and would align ACM with the emerging norm of retraction transparency.
}%

\subsection{Implications for Retraction Practice}%
\label{sec:disc:implications}%
\notouch{%
Authors, reviewers, program committees, and funding bodies have a stake in the integrity of all ten publishers examined here.
IEEE's experience since 2019 demonstrates that large-scale retraction is operationally feasible, with the shortest turnaround times in the dataset (Section~\ref{sec:disc:landscape}).
Proactive retraction need not be protracted or impractical.
}%

\notouch{%
If retraction at scale is operationally feasible, the question shifts to publishers whose records suggest underuse of the mechanism.
The findings of this study indicate that ACM is one such case.
Two concrete steps would strengthen confidence in the ACM Digital Library.
First, applying the COPE guidelines it has adopted for a subset of its journals~\cite{retraction-guidelines-cope.pdf} across the full ACM Digital Library would ensure that problematic work is corrected in a manner consistent with the commitments ACM has already made.
The other COPE-member publishers in this study (Elsevier, Springer Nature, Wiley, Taylor~\&~Francis, SAGE, and PLoS) operate at this threshold.
This need not mean abandoning careful deliberation; the COPE framework itself calls for evidence-based adjudication.
It would mean lowering the bar from ``extremely high'' to the evidence-based standard ACM's own COPE membership prescribes.
Second, making the dark archive's metadata publicly accessible would enable the research community to assess the full scope of corrections and update citation records accordingly.
}%

\notouch{%
We acknowledge that lowering a retraction threshold carries risks.
False positives -- retracting work that is subsequently vindicated -- can cause reputational harm and erode trust in the process itself.
The Retraction Watch data include 98 reinstatements (87.8\% from Elsevier; Table~\ref{tab:retraction-nature}), attesting to the fact that retraction decisions are not always final.
But the status quo also carries costs.
When flawed research circulates uncorrected, downstream studies build on unreliable foundations~\cite{e031909.full.pdf, pone.0285383.pdf}, misinformation spreads~\cite{3637335.pdf}, and public trust in the scientific enterprise is undermined.
The balance of these risks -- overcorrection versus undercorrection -- is a judgment each publisher must make.
The data suggest that ACM's current position on this spectrum is an outlier, and that the question of whether recalibration is warranted deserves open discussion.%
}%

\notouch{%
The broader publishing ecosystem is already moving toward greater transparency.
The integration of Retraction Watch into Crossref~\cite{Crossref}, the development of retraction-aware retrieval \cite{3576840.3578281} and detection \cite{paper-mill-detector,Scancare087581} systems, and the growing body of research on the consequences of retracted publications~\cite{qss_a_00303.pdf, qss_a_00243.pdf} all point in the same direction: openness, accountability, and prompt correction.
ACM's current practices are out of step with this trajectory and with its own commitments as a COPE member.
Aligning with these norms would strengthen the ACM Digital Library and signal that research integrity is a priority commensurate with the importance of the work ACM publishes.
}%


\section{Conclusion}%
\label{sec:conclusion}%
\notouch{%
This study presented a large-scale bibliometric analysis of 46,087 retractions across ten publishers, drawing on the Retraction Watch database for the period 1997--2026.
We examined retraction patterns across nine dimensions including normalized retraction rates, retraction reasons, temporal trends, and geographic distributions, comparing ACM in detail against the dataset's other nine publishers.
}%

\notouch{%
The cross-publisher comparison reveals that retraction practices vary across every dimension we measured, with no single publisher profile constituting a normative baseline.
Within this variation, ACM's record stands apart: the most concentrated reason profile in the dataset, with seven of the ten most common global retraction reasons entirely absent, and a temporal record that begins only in 2020 despite a catalog dating to 1997.
The most parsimonious explanation is that ACM's self-described ``extremely high'' retraction threshold 
stops the publisher to act on 
the types of cases that other publishers routinely retract.
The non-public dark archive compounds this concern by preventing independent verification.
Two steps would strengthen confidence in the ACM Digital Library: aligning retraction practice with its (partial) COPE commitment, and making the dark archive's metadata publicly accessible.%
}%

\notouch{%
More broadly, the integration of the Retraction Watch database into Crossref~\cite{Crossref} is creating infrastructure for retraction-aware citation analysis and integrity monitoring across all publishers. As this infrastructure matures, the patterns documented here will become easier to track over time.
This work is intended to inform a constructive discussion about how retraction practices differ across publishers and what those differences mean for the integrity of the scholarly record.
}%


\section*{Author Contributions}
JO: Conceptualization, Data curation, Formal analysis, Investigation, Methodology, Project administration, Software, Validation, Visualization, Writing -- original draft.

\section*{Competing Interests}
The author declares no competing interests.

\section*{Funding}
This research received no specific grant from any funding agency.

\section*{Data Availability}
The Retraction Watch database is publicly available at \url{http://retractiondatabase.org}.
Raw bibliometric data from Dimensions (\url{https://www.dimensions.ai}) is subject to license restrictions and cannot be shared directly. Analysis scripts
are available at \url{https://github.com/retractions/retractions.github.io}.
Researchers with 
access to Dimensions can reproduce the queries described in the Method section.
{
\small
\doublespacing
\raggedright
\bibliographystyle{apacite}
\bibliography{paper}

@article{paper-mill-challenges,
title={Paper mill challenges: past, present, and future},
author={
Parker, Lisa and
Boughton, Stephanie and
Bero, Lisa and
Byrne, Jennifer A.
},
journal={Journal of Clinical Epidemiology},
volume={176},
number={},
year={2024},
doi={10.1016/j.jclinepi.2024.111549},
pages={},
publisher={Elsevier},
}

@article{Scancare087581,
	author = {Scancar, Baptiste and Byrne, Jennifer A and Causeur, David and Barnett, Adrian G},
	title = {Machine learning based screening of potential paper mill publications in cancer research: methodological and cross sectional study},
	volume = {392},
	elocation-id = {e087581},
	year = {2026},
	doi = {10.1136/bmj-2025-087581},
	publisher = {BMJ Publishing Group Ltd},
	journal = {BMJ},
}

@article{paper-mill-detector,
author = {Holly Else},
title = {Paper-mill detector tested in push to stamp out fake science},
journal = {Nature},
volume = {612},
pages = {386--387},
year = {2022},
doi = {},
}

@article{science.342.6162.1035,
author = {Mara Hvistendahl},
title = {China's Publication Bazaar},
journal = {Science},
volume = {342},
number = {6162},
pages = {1035-1039},
year = {2013},
doi = {10.1126/science.342.6162.1035},
}

@article{ferguson2014peer-review-scam,
  author  = {Cat Ferguson and Adam Marcus and Ivan Oransky},
  title   = {Publishing: The peer-review scam},
  journal = {Nature},
  volume  = {515},
  pages   = {480--482},
  year    = {2014},
  doi     = {10.1038/515480a},
}

@misc{kincaid2024iospress,
    author = {Ellie Kincaid},
    title  = {Exclusive: Publisher retracts more than 450 papers from journal it acquired last year},
    year   = {2024},
    url    = {https://retractionwatch.com/2024/08/23/exclusive-publisher-retracts-more-than-450-papers-from-journal-it-acquired-last-year/},
  }

@article{ACMDL-launch,
author = {Denning, Peter J.},
title = {The ACM digital library goes live},
year = {1997},
issue_date = {July 1997},
publisher = {Association for Computing Machinery},
address = {New York, NY, USA},
volume = {40},
number = {7},
issn = {0001-0782},
doi = {10.1145/256175.256179},
journal = {Commun. ACM},
month = jul,
pages = {28–29},
numpages = {2}
}

@article{elife-02956-v1.pdf,
title = {Financial costs and personal consequences of research misconduct resulting in retracted publications},
author = {Stern, Andrew M and Casadevall, Arturo and Steen, R Grant and Fang, Ferric C},
editor = {Rodgers, Peter},
volume = 3,
year = 2014,
month = {aug},
pub_date = {2014-08-14},
pages = {e02956},
citation = {eLife 2014;3:e02956},
doi = {10.7554/eLife.02956},
journal = {eLife},
issn = {2050-084X},
publisher = {eLife Sciences Publications, Ltd},
}

@article{Research-misconduct-in-hospitals-is-spreading-A-bibliometric-analysis-of-retracted-papers-from-Chinese-universityaffiliated-hospitals.pdf,
author = {Zi-han Yuan and Yi Liu},
doi = {10.2478/jdis-2023-0022},
title = {Research misconduct in hospitals is spreading: A bibliometric analysis of retracted papers from {Chinese} university-affiliated hospitals},
journal = {Journal of Data and Information Science},
number = {4},
volume = {8},
year = {2023},
pages = {84--101}
}

@misc{2406.07016.pdf,
      title={Delving into {ChatGPT} usage in academic writing through excess vocabulary}, 
      author={Dmitry Kobak and Rita González-Márquez and Emőke-Ágnes Horvát and Jan Lause},
      year={2024},
      eprint={2406.07016},
      archivePrefix={arXiv},
}

@misc{gino,
title={In Historic Step, Harvard Moves Toward Tenure Revocation for Business School Professor Gino},
url={https://www.thecrimson.com/article/2023/8/28/gino-tenure-review-begins/},
year={2023},
publisher={The Harvard Crimson},
author={Rahem D. Hamid and Neil H. Shah and Claire Yuan},
}

@inproceedings{zlabinger2017.pdf,
author = {Zlabinger, Markus and Hanbury, Allan},
title = {Finding duplicate images in biology papers},
year = {2017},
isbn = {9781450344869},
publisher = {Association for Computing Machinery},
address = {New York, NY, USA},
doi = {10.1145/3019612.3019875},
booktitle = {Proceedings of the Symposium on Applied Computing},
pages = {957–959},
numpages = {3},
location = {Marrakech, Morocco},
series = {SAC '17}
}

@article{FEBS,
author = {Christopher, Jana},
title = {Systematic fabrication of scientific images revealed},
journal = {FEBS Letters},
volume = {592},
number = {18},
pages = {3027-3029},
doi = {https://doi.org/10.1002/1873-3468.13201},
year = {2018}
}

@article{edwards-roy-2017-academic-research-in-the-21st-century-maintaining-scientific-integrity-in-a-climate-of-perverse.pdf,
author = {Edwards, Marc A. and Roy, Siddhartha},
title = {Academic Research in the 21st Century: Maintaining Scientific Integrity in a Climate of Perverse Incentives and Hypercompetition},
journal = {Environmental Engineering Science},
volume = {34},
number = {1},
pages = {51-61},
year = {2017},
doi = {10.1089/ees.2016.0223},
}

@article{tijdink2014.pdf,
author = {Joeri K. Tijdink and Reinout Verbeke and Yvo M. Smulders},
title ={Publication Pressure and Scientific Misconduct in Medical Scientists},
journal = {Journal of Empirical Research on Human Research Ethics},
volume = {9},
number = {5},
pages = {64-71},
year = {2014},
doi = {10.1177/1556264614552421},
}

@article{haider_gpt_fabricated_scientific_papers_20240903.pdf,
title={{GPT}-fabricated  scientific  papers  on  {Google Scholar}:  Key features, spread, and implications for preempting evidence 
manipulation},
author={Jutta Haider and Kristofer Rolf Söderström and Björn Ekström and Malte Rödl},
year={2024},
journal={Harvard Kennedy School (HKS) Misinformation Review},
volume={5},
number={5},
}

@article{s11192-022-04321-w.pdf,
title={Citation of retracted research: a case-controlled, ten-year follow-up scientometric analysis of {Scott S. Reuben}'s malpractice},
author={
    Szilagyi, Istvan-Szilard and
Schittek, Gregor A. and
Klivinyi, Christoph and
Simonis, Holger and
Ulrich, Torsten and
Bornemann-Cimenti, Helmar
},
year={2022},
journal={Scientometrics},
volume={127},
number={5},
doi={10.1007/s11192-022-04321-w},
pages={2611-2620},
}

@article{el-menyar-et-al-2021-publications-and-retracted-articles-of-covid-19-pharmacotherapy-related-research-a-systematic.pdf,
author = {Ayman El-Menyar and Ahammed Mekkodathil and Mohammad Asim and Rafael Consunji and Sandro Rizoli and Ahmed Abdel-Aziz Bahey and Hassan Al-Thani},
title ={Publications and retracted articles of {COVID-19} pharmacotherapy-related research: A systematic review},

journal = {Science Progress},
volume = {104},
number = {2},
pages = {00368504211016936},
year = {2021},
doi = {10.1177/00368504211016936},
}

@article{e031909.full.pdf,
	author = {Avenell, Alison and Stewart, Fiona and Grey, Andrew and Gamble, Greg and Bolland, Mark},
	title = {An investigation into the impact and implications of published papers from retracted research: systematic search of affected literature},
	volume = {9},
	number = {10},
	elocation-id = {e031909},
	year = {2019},
	doi = {10.1136/bmjopen-2019-031909},
	publisher = {British Medical Journal Publishing Group},
	issn = {2044-6055},
	journal = {BMJ Open}
}

@article{pone.0285383.pdf,
title={An analysis of retracted papers in Computer Science},
doi={10.1371/journal.pone.0285383},
author={Shepperd, Martin and Yousefi, Leila},
journal={PLoS One},
year={2023},
volume={18},
number={5},
}

@inproceedings{yang2020.pdf,
author = {Yang, Siluo and Qi, Fan and Diao, Heyu},
title = {Exploring the Influence of Publication Retraction on Citations in Psychology Science},
year = {2020},
isbn = {9781450375856},
publisher = {Association for Computing Machinery},
address = {New York, NY, USA},
doi = {10.1145/3383583.3398583},
booktitle = {Proceedings of the {ACM/IEEE} Joint Conference on Digital Libraries in 2020},
pages = {501–502},
numpages = {2},
series = {JCDL '20}
}

@misc{retractionwatch,
title={{Retraction Watch Database}},
author={{Retraction Watch}},
year={2024},
url={http://retractiondatabase.org},
}

@misc{ACM-POLICY,
title={{ACM} Publications Policy on the Withdrawal, Correction, Retraction, and Removal of Works from {ACM Publications} and {ACM DL}},
author={{ACM}},
year={2018},
url={https://www.acm.org/publications/policies/retraction-policy},
}

@misc{Crossref,
title={Crossref acquires {Retraction Watch} data and opens it for the scientific community},
author={Ginny Hendricks and Rachael Lammey and Lucy Ofiesh and Geoffrey Bilder and Ed Pentz},
year={2023},
doi={10.13003/c23rw1d9},
}

@misc{more-than-300-at-once,
title={More than 300 at once: Publisher retracts entire conference proceedings},
author={Ivan Oransky},
year={2022},
url={https://retractionwatch.com/2022/04/20/more-than-300-at-once-publisher-retracts-entire-conference-proceedings/},
}

@article{3637335.pdf,
author = {Yaqub, Waheeb and Kay, Judy and Goldwater, Micah},
title = {Foundations for Enabling People to Recognise Misinformation in Social Media News based on Retracted Science},
year = {2024},
issue_date = {April 2024},
publisher = {Association for Computing Machinery},
address = {New York, NY, USA},
volume = {8},
number = {CSCW1},
doi = {10.1145/3637335},
journal = {Proc. ACM Hum.-Comput. Interact.},
month = {apr},
articleno = {58},
numpages = {38}
}

@misc{disturbing-experience,
title={`{A} disturbing experience': Postdoc fights to have work that plagiarized her thesis retracted},
author={Dawn Attride},
year={2024},
url={https://retractionwatch.com/2024/06/03/a-disturbing-experience-postdoc-fights-to-have-work-that-plagiarized-her-thesis-retracted/},
}

@article{science.aao0185.pdf,
author = {Santo Fortunato  and Carl T. Bergstrom  and Katy Börner  and James A. Evans  and Dirk Helbing  and Staša Milojević  and Alexander M. Petersen  and Filippo Radicchi  and Roberta Sinatra  and Brian Uzzi  and Alessandro Vespignani  and Ludo Waltman  and Dashun Wang  and Albert-László Barabási },
title = {Science of science},
journal = {Science},
volume = {359},
number = {6379},
pages = {eaao0185},
year = {2018},
doi = {10.1126/science.aao0185},
}

@article{MetaResearch,
  author = {Ioannidis, John P. A. and Fanelli, Daniele and Dunne, Debbie Drake and Goodman, Steven N.},
  title = {Meta-research: Evaluation and Improvement of Research Methods and Practices},
  journal = {PLoS Biology},
  year = {2015},
  month = {Oct},
  day = {2},
  volume = {13},
  number = {10},
  doi = {10.1371/journal.pbio.1002264},
}

@misc{2107.06751.pdf,
      title={Tortured phrases: A dubious writing style emerging in science. {Evidence} of critical issues affecting established journals}, 
      author={Guillaume Cabanac and Cyril Labbé and Alexander Magazinov},
      year={2021},
      eprint={2107.06751},
      archivePrefix={arXiv},
}

@misc{retractions-from-xplore.html,
title={{IEEE} Announces Retractions from the {IEEE Xplore® Digital Library}},
year={2019},
author={Monika Stickel and Francine Tardo},
url={https://www.ieee.org/about/news/2019/retractions-from-xplore.html},
}

@inproceedings{3576840.3578281,
author = {Wang, Peiling},
title = {Rising of Retracted Research Works and Challenges in Information Systems: Need New Features for Information Retrieval and Interactions},
year = {2023},
isbn = {9798400700354},
publisher = {Association for Computing Machinery},
address = {New York, NY, USA},
doi = {10.1145/3576840.3578281},
booktitle = {Proceedings of the 2023 Conference on Human Information Interaction and Retrieval},
pages = {69–82},
numpages = {14},
series = {CHIIR '23}
}

@article{qss_a_00303.pdf,
    author = {Woo, Seokkyun and Walsh, John P.},
    title = "{On the shoulders of fallen giants: What do references to retracted research tell us about citation behaviors?}",
    journal = {Quantitative Science Studies},
    volume = {5},
    number = {1},
    pages = {1-30},
    year = {2024},
    month = {03},
    issn = {2641-3337},
    doi = {10.1162/qss_a_00303},
}

@article{qss_a_00222.pdf,
    author = {Heibi, Ivan and Peroni, Silvio},
    title = "{A quantitative and qualitative open citation analysis of retracted articles in the humanities}",
    journal = {Quantitative Science Studies},
    volume = {3},
    number = {4},
    pages = {953-975},
    year = {2022},
    month = {12},
    issn = {2641-3337},
    doi = {10.1162/qss_a_00222},
}

@article{qss_a_00243.pdf,
    author = {Malkov, Dmitry and Yaqub, Ohid and Siepel, Josh},
    title = "{The spread of retracted research into policy literature}",
    journal = {Quantitative Science Studies},
    volume = {4},
    number = {1},
    pages = {68-90},
    year = {2023},
    month = {03},
    issn = {2641-3337},
    doi = {10.1162/qss_a_00243},
}

@article{else2021.pdf,
volume = {591},
author = {Else, Holly and Van Noorden, Richard},
address = {Berlin},
issn = {0028-0836},
journal = {Nature},
language = {eng},
number = {7851},
pages = {516-519},
publisher = {Springer},
doi={10.1038/d41586-021-00733-5},
title = {The Battle Against Paper Mills},
year = {2021},
}

@misc{Elite,
url={https://www.nature.com/articles/d41586-024-01697-y},
author={Smriti Mallapaty},
title={Elite researchers in China say they had `no choice' but to commit misconduct},
year={2024},
}

@article{zhang-wang-2024-research-misconduct-in-china-towards-an-institutional-analysis.pdf,
author = {Xinqu Zhang and Peng Wang},
title ={Research misconduct in {China}: towards an institutional analysis},
journal = {Research Ethics},
volume = {21},
number = {1},
pages = {76-96},
year = {2025},
doi = {10.1177/17470161241247720},
}

@article{cetina2007.pdf,
author = {Karin Knorr Cetina},
title = {Culture in global knowledge societies: knowledge cultures and epistemic cultures},
journal = {Interdisciplinary Science Reviews},
volume = {32},
number = {4},
pages = {361--375},
year = {2007},
publisher = {Taylor \& Francis},
doi = {10.1179/030801807X163571},
}

@misc{Oransky,
title={There's far more scientific fraud than anyone wants to admit},
author={Ivan Oransky and Adam Marcus},
year={2023},
url={https://www.theguardian.com/commentisfree/2023/aug/09/scientific-misconduct-retraction-watch},
publisher={The Guardian},
}

@article{d41586-023-03974-8.pdf,
title={More than 10,000 Research Papers were Retracted in 2023 -- A New Record},
author={Richard Van Noorden},
year={2023},
journal={Nature},
volume={624},
number={},
pages={479-481},
publisher={Springer},
}

@misc{retraction-guidelines-cope.pdf,
title={COPE Guidelines: Retraction Guidelines},
author={{COPE Council}},
year={2019},
doi={10.24318/cope.2019.1.4},
}

@ARTICLE{Hook2018Dimensions,    
AUTHOR={Hook, Daniel W.  and Porter, Simon J.  and Herzog, Christian },
TITLE={Dimensions: Building Context for Search and Evaluation},
JOURNAL={Frontiers in Research Metrics and Analytics},
VOLUME={3},
YEAR={2018},
DOI={10.3389/frma.2018.00023},
ISSN={2504-0537},
}
}

\appendix


\newpage

\section{Retraction Lag Statistics}
\label{appendix:lag}

\begin{table}[htbp]
	\centering
	\caption{Retraction lag statistics by publisher (in days), sorted by median lag.}
	\label{tab:retraction-lag}
	\small
	\begin{tabular}{l r r r r r r r r}
		\toprule
		Publisher & N & Mean & Median & SD & Q1 & Q3 & Min & Max \\
		\midrule
		PLoS & 1,080 & 1,825 & 1,568 & 1,433 & 512 & 2,865 & 2 & 6,039 \\
		Wiley & 3,817 & 1,465 & 940 & 1,561 & 409 & 1,938 & 0 & 11,032 \\
		SAGE & 1,217 & 1,142 & 818 & 1,496 & 422 & 1,217 & 0 & 12,563 \\
		Taylor \& Francis & 2,314 & 982 & 758 & 1,019 & 266 & 1,267 & 0 & 10,964 \\
		IOS Press & 1,671 & 908 & 651 & 796 & 386 & 1,092 & 0 & 8,246 \\
		Springer Nature & 7,534 & 1,081 & 643 & 1,365 & 264 & 1,400 & 0 & 16,626 \\
		Elsevier & 6,520 & 1,100 & 558 & 1,631 & 139 & 1,368 & 0 & 16,950 \\
		Hindawi & 11,486 & 541 & 480 & 320 & 378 & 621 & 0 & 6,891 \\
		ACM & 354 & 200 & 189 & 110 & 189 & 189 & 2 & 1,256 \\
		IEEE & 10,094 & 64 & 41 & 244 & 29 & 60 & 0 & 7,915 \\
		\bottomrule
	\end{tabular}
\end{table}

\section{Consolidated Retraction Reasons}%
\label{sec:retractions}%

%
%


We consolidated the 111 original Retraction Watch reasons into a set of broader categories (see Section~\ref{sec:method}).
In this section, we describe the consolidated reasons and provide a mapping to Retraction Watch (see Table~\ref{tab:reason-consolidation}).


\paragraph{Peer Review}
This category consolidates reason codes related to compromised or manipulated peer review processes, including ``Fake Peer Review'' and ``Taken via Peer Review,'' as well as the high-frequency pass-through reason code ``Compromised Peer Review.''

\paragraph{Data}
This category encompasses ``Concerns/Issues About Data,'' ``Falsification/Fabrication of Data,'' ``Original Data not Provided,'' and ``Error in Data.''

\paragraph{Authorship}
Authorship-related reasons include ``Concerns/Issues About Authorship,'' ``False/Forged Authorship,'' and ``Conflict of Interest.''

\paragraph{Results}
This category consolidates ``Concerns/Issues About Results,'' ``Unreliable Results,'' and ``Results Not Reproducible.''

\paragraph{Methods}
Methodological concerns are captured through ``Error in Analyses'' and ``Error in Methods.''

\paragraph{Duplication}
This category maps ``Duplication of Article'' and ``Euphemisms for Duplication,'' covering cases where substantially identical content was published in multiple venues.

\paragraph{Third Party}
Third-party-related reasons include ``Investigation by Third Party,'' ``Investigation by Company/Institution,'' ``Civil Proceedings,'' ``Objections by Third Party,'' and ``Concerns/Issues about Third Party Involvement.''

\paragraph{Text}
This category consolidates ``Randomly Generated Content'' and ``Error in Text,'' covering cases where the textual content of an article was identified as machine-generated or contained substantive textual errors.


\paragraph{Information}
This meta-category consolidates ``Notice -- Limited or No Information'' and ``Notice -- No/Limited Information,'' representing retraction notices that provide minimal or no explanation for the retraction.

\paragraph{Misconduct}
Misconduct-related reason codes include ``Breach of Policy by Author,'' ``Misconduct by Author,'' ``Misconduct by Company/Institution,'' and ``Criminal Proceedings.''

\paragraph{Plagiarism}
This category consolidates seven original reason codes: ``Euphemisms for Plagiarism,'' ``Plagiarism of Text,'' ``Plagiarism of Article,'' ``Plagiarism of Image,'' ``Referencing/Attribution,'' ``Taken from Dissertation/Thesis,'' and ``Copyright Claims.''

\paragraph{Notice}
This category captures ``Notice -- Lack of,'' referring to entries where no formal retraction notice was issued or could be located.


\paragraph{Author}
This category consolidates ``Lack of Approval from Author'' and ``Objections by Author(s),'' representing cases where the author did not consent to the publication or raised objections post-publication.

\paragraph{Approval}
Approval-related reasons include ``Lack of Approval from Third Party'' and ``Lack of IRB/IACUC Approval,'' capturing retractions due to missing institutional or ethical review board approvals.

\paragraph{Materials}
This category covers ``Contamination of Materials (General),'' referring to cases where physical or biological materials used in the research were compromised.
ACM has 0 entries compared to 2 for IEEE.

\paragraph{Retract and Replace}
This category retains the original Retraction Watch reason code ``Retract and Replace'' without consolidation. It refers to cases where an article is simultaneously retracted and replaced with a corrected version, rather than being corrected via an erratum or corrigendum.

\paragraph{Other}
The Other category consolidates ``Date of Retraction/Other Unknown'' and ``Doing the Right Thing,'' as well as the high-frequency pass-through reason code ``Date of Article and/or Notice Unknown.''


\begin{table*}[htb]
	\caption{Consolidation mapping from original Retraction Watch reasons to our consolidated categories. Each row shows the target category and the original reasons that are mapped to it via substring matching.
		Reasons not listed here pass through unchanged.
	}
	\label{tab:reason-consolidation}
	\centering
	\small
	\begin{tabular}{lp{10.5cm}}
		\toprule
		Consolidated Category & Retraction Watch Reason(s) \\
		\midrule
		Approval & Lack of Approval from Third Party; Lack of IRB/IACUC Approval \\
		\addlinespace
		Author & Lack of Approval from Author; Objections by Author(s) \\
		\addlinespace
		Authorship & Concerns/Issues About Authorship; False/Forged Authorship; Conflict of Interest \\
		\addlinespace
		Data & Concerns/Issues About Data; Falsification/Fabrication of Data; Original Data not Provided; Error in Data \\
		\addlinespace
		Duplication & Duplication of Article; Euphemisms for Duplication \\
		\addlinespace
		Information & Notice -- Limited or No Information; Notice -- No/Limited Information \\
		\addlinespace
		Materials & Contamination of Materials (General) \\
		\addlinespace
		Methods & Error in Analyses; Error in Methods \\
		\addlinespace
		Misconduct & Breach of Policy by Author; Misconduct by Author; Misconduct by Company/Institution; Criminal Proceedings \\
		\addlinespace
		Notice & Notice -- Lack of \\
		\addlinespace
		Other & Date of Retraction/Other Unknown; Doing the Right Thing \\
		\addlinespace
		Peer Review & Fake Peer Review; Taken via Peer Review \\
		\addlinespace
		Plagiarism & Euphemisms for Plagiarism; Plagiarism of Text; Referencing/Attribution; Plagiarism of Article; Plagiarism of Image; Taken from Dissertation/Thesis; Copyright Claims \\
		\addlinespace
		Results & Concerns/Issues About Results; Unreliable Results; Results Not Reproducible \\
		\addlinespace
		Retract and Replace & Retract and Replace \\
		\addlinespace
		Text & Randomly Generated Content; Error in Text \\
		\addlinespace
		Third Party & Investigation by Third Party; Investigation by Company/Institution; Civil Proceedings; Objections by Third Party; Concerns/Issues about Third Party Involvement \\
		\bottomrule
	\end{tabular}
\end{table*}


\begin{table*}[htbp]
  \centering
  \caption{Sensitivity analysis of reason consolidation. Three alternative mappings (S1--S3) are compared against the baseline (S0). S1 moves Conflict of Interest from Authorship to Misconduct. S2 moves Copyright Claims and Taken from Dissertation/Thesis from Plagiarism to Other. S3 combines all contested reassignments (S1 + S2 plus Criminal Proceedings to Third Party, Referencing/Attribution to Authorship, and Breach of Policy by Author to Other). Max $|\Delta|$ is the largest absolute percentage-point shift for any reason category. $\rho$ is the Spearman rank correlation between baseline and alternative count vectors (computed over non-zero categories only). Top-3 indicates whether the three most frequent reasons match the baseline in identity and rank order (\checkmark = stable, $\times$ = changed).}
  \label{tab:sensitivity-reasons}
  \small
  \begin{tabular}{l r rrr rrr rrr}
    \toprule
    & & \multicolumn{3}{c}{S1} & \multicolumn{3}{c}{S2} & \multicolumn{3}{c}{S3} \\
    \cmidrule(lr){3-5} \cmidrule(lr){6-8} \cmidrule(lr){9-11}
    Publisher & N & $|\Delta|$ & $\rho$ & Top-3 & $|\Delta|$ & $\rho$ & Top-3 & $|\Delta|$ & $\rho$ & Top-3 \\
    \midrule
    ACM & 354 & 0.28 & 0.500 & \checkmark & 0.00 & 1.000 & \checkmark & 0.28 & 0.500 & \checkmark \\
    IEEE & 10,094 & 0.00 & 1.000 & \checkmark & 0.13 & 0.993 & \checkmark & 43.46 & 0.941 & $\times$ \\
    Elsevier & 6,520 & 2.87 & 0.999 & \checkmark & 2.44 & 0.995 & \checkmark & 5.51 & 0.988 & \checkmark \\
    Springer Nature & 7,534 & 0.40 & 1.000 & \checkmark & 2.43 & 0.984 & $\times$ & 21.81 & 0.975 & $\times$ \\
    Wiley & 3,817 & 1.07 & 1.000 & \checkmark & 1.02 & 0.996 & \checkmark & 11.06 & 0.986 & \checkmark \\
    Taylor \& Francis & 2,314 & 0.13 & 1.000 & \checkmark & 1.60 & 0.988 & \checkmark & 3.41 & 0.977 & \checkmark \\
    SAGE & 1,217 & 1.73 & 0.996 & \checkmark & 0.66 & 0.998 & \checkmark & 19.88 & 0.992 & \checkmark \\
    Hindawi & 11,486 & 0.05 & 0.999 & \checkmark & 0.18 & 0.969 & \checkmark & 73.32 & 0.940 & \checkmark \\
    IOS Press & 1,671 & 0.06 & 0.999 & \checkmark & 0.06 & 0.998 & \checkmark & 66.73 & 0.942 & $\times$ \\
    PLoS & 1,080 & 15.74 & 0.976 & \checkmark & 4.07 & 0.978 & \checkmark & 14.81 & 0.963 & \checkmark \\
    \bottomrule
  \end{tabular}
\end{table*}

\section{Author Disambiguation Sensitivity Analysis}
\label{appendix:sensitivity-authors}

To assess the robustness of our repeat-author findings, we compared three disambiguation strategies: name-only (S1), name+institution with Dimensions \texttt{researcher\_id} (S2, used in the main analysis), and Dimensions \texttt{researcher\_id} only (S3).
Table~\ref{tab:sensitivity-authors} shows that S2 and S3 converge on the same count of authors with 50 or more retractions (22), while the name-only baseline (S1) inflates this to 68 due to name collisions.
Per-publisher comparisons (Table~\ref{tab:sensitivity-authors-per-publisher}) confirm that repeat-author rates are stable across strategies, with the largest differences appearing in publishers whose author pools contain many common names.


\begin{table}[htbp]
  \centering
  \caption{Sensitivity analysis of author disambiguation strategies. S1 uses author name only, S2 uses Dimensions \texttt{researcher\_id} with name+institution fallback (primary analysis), and S3 uses Dimensions \texttt{researcher\_id} only.}
  \label{tab:sensitivity-authors}
  \small
  \begin{tabular}{l r r r}
    \toprule
    Metric & S1 (Name) & S2 (Name+Inst.) & S3 (Dim. only) \\
    \midrule
    Unique authors & 121,254 & 148,508 & 111,805 \\
    With 2+ retractions & 21,868 & 15,608 & 15,086 \\
    With 5+ retractions & 3,376 & 1,482 & 1,478 \\
    With 10+ retractions & 961 & 331 & 331 \\
    With 50+ retractions & 51 & 13 & 13 \\
    \bottomrule
  \end{tabular}
\end{table}

\begin{table*}[htbp]
  \centering
  \caption{Per-publisher unique author counts and repeat-author rates (2+) under each disambiguation strategy.}
  \label{tab:sensitivity-authors-per-publisher}
  \small
  \begin{tabular}{l r r r r r r}
    \toprule
    & \multicolumn{2}{c}{S1 (Name)} & \multicolumn{2}{c}{S2 (Name+Inst.)} & \multicolumn{2}{c}{S3 (Dim. only)} \\
    \cmidrule(lr){2-3} \cmidrule(lr){4-5} \cmidrule(lr){6-7}
    Publisher & Authors & 2+ (\%) & Authors & 2+ (\%) & Authors & 2+ (\%) \\
    \midrule
    ACM & 621 & 5.8 & 626 & 5.1 & 218 & 11.0 \\
    IEEE & 20,333 & 16.8 & 24,112 & 7.7 & 9,776 & 16.5 \\
    Elsevier & 25,483 & 11.0 & 26,263 & 9.1 & 24,513 & 9.6 \\
    Springer Nature & 26,654 & 11.8 & 28,277 & 8.5 & 26,479 & 9.0 \\
    Wiley & 14,463 & 14.2 & 14,965 & 11.4 & 12,172 & 14.0 \\
    Taylor \& Francis & 7,828 & 8.8 & 8,224 & 5.6 & 6,050 & 7.2 \\
    SAGE & 3,854 & 7.3 & 3,888 & 5.7 & 3,084 & 7.2 \\
    Hindawi & 30,383 & 17.4 & 37,781 & 7.2 & 27,069 & 9.7 \\
    IOS Press & 4,549 & 10.6 & 4,715 & 7.9 & 3,283 & 11.1 \\
    PLoS & 6,192 & 11.2 & 6,321 & 9.8 & 5,735 & 10.8 \\
    \bottomrule
  \end{tabular}
\end{table*}

\section{Dimensions--Crossref Cross-Validation}
\label{appendix:crossref-validation}

To assess whether Dimensions publication counts are consistent across publishers, we queried the Crossref REST API\footnote{\url{https://api.crossref.org}} for each publisher's member record, filtering to journal articles and proceedings articles over the same period (1997--2026).
IOS Press additionally includes book chapters, consistent with the Dimensions approach described in Section~\ref{sec:normalization}.
Table~\ref{tab:crossref-validation} reports per-publisher deviations.
The two sources agree within 9\% for nine of ten publishers.
Hindawi is the exception: its Crossref count (113,040) is 68.5\% lower than the Dimensions total (358,879) because Wiley's January 2023 acquisition reassigned many Hindawi DOIs to Wiley's Crossref member record.

%

\begin{table}[htb]
\caption{Cross-validation of Dimensions publication counts against the Crossref REST API (1997--2026). 
Both sources are filtered to journal articles and proceedings articles (IOS Press additionally includes book chapters). 
Pearson~$r = 0.9985$; mean absolute percentage deviation = 9.7\%.}
\label{tab:crossref-validation}
\centering
\small
\begin{tabular}{lrrr}
\toprule
Publisher & Dimensions & Crossref & Deviation (\%) \\
\midrule
Elsevier             &   15,686,751 &   17,076,960 &   +8.9 \\
Springer Nature      &    7,994,821 &    8,042,203 &   +0.6 \\
Wiley                &    6,071,179 &    6,594,454 &   +8.6 \\
IEEE                 &    5,703,250 &    5,373,462 &   -5.8 \\
Taylor \& Francis    &    3,534,113 &    3,572,909 &   +1.1 \\
SAGE                 &    2,179,130 &    2,209,788 &   +1.4 \\
ACM                  &      626,110 &      636,854 &   +1.7 \\
PLoS                 &      401,947 &      402,443 &   +0.1 \\
Hindawi              &      358,879 &      113,040 &  -68.5 \\
IOS Press            &       57,794 &       57,820 &   +0.0 \\
\bottomrule
\end{tabular}
\end{table}


\end{document}